\documentclass[prd,superscriptaddress,twocolumn,showpacs,amsmath,amssymb,floatfix,nofootinbib]{revtex4}
\usepackage{graphicx}

\newcommand{\beqn}{\begin{eqnarray}}
\newcommand{\eeqn}{\end{eqnarray}}
\newcommand{\eq}[1]{(\ref{#1})}

\newcommand{\dD}{{\mathrm D}}

\newcommand{\cZ}{{\mathcal Z}}

\newcommand{\dd}{{\mathrm d}}
\newcommand{\Tr}{{\mathrm{Tr}}\,}
\newcommand{\Z}{{\mathbb Z}}
\newcommand{\vort}{{\mathrm{vort}}}
\newcommand{\novort}{{\mathrm{novort}}}

\def\bbbone{{\mathchoice {\rm 1\mskip-4mu l} {\rm 1\mskip-4mu l}
{\rm 1\mskip-4.5mu l} {\rm 1\mskip-5mu l}}}

\begin{document}

\title{Manifestations of magnetic vortices in the equation of state of a Yang-Mills plasma}

\author{M.N.~Chernodub}
\affiliation{Institute of Theoretical and Experimental Physics, B.Cheremushkinskaya 25, Moscow, 117218, Russia}
\author{Atsushi Nakamura}
\affiliation{Research Institute for Information Science and Education, Hiroshima University, Higashi-Hiroshima, 739-8527, Japan}
\author{V.I.~Zakharov}
\affiliation{Istituto Nazionale di Fisica Nucleare -- Sezione di Pisa,
Dipartimento di Fisica Universita di Pisa, Largo Pontecorvo 3, 56127 Pisa, Italy}
\affiliation{Institute of Theoretical and Experimental Physics, B.Cheremushkinskaya 25, Moscow, 117218, Russia}

\begin{abstract}
The vacuum of Yang-Mills theory contains singular stringlike objects identified
with center (magnetic) vortices. Percolation of magnetic vortices is
known to be responsible for the color confinement in the low-temperature
phase of the theory. In our work we study properties of the vortices at finite
temperature using lattice simulations of $SU(2)$ gauge theory. We show
that magnetic vortices provide a numerically large contribution to
thermodynamic quantities of the gluon plasma in Yang-Mills theory.
In particular, we observe that in the deconfinement phase at temperatures
$T_c <  T \lesssim 3 T_c$ the magnetic component of the gluon plasma produces
a negative (ghostlike) contribution to the anomaly of the energy-momentum tensor.
In the confinement phase the vortex contribution is positive.
The thermodynamical significance of the magnetic objects allows us to suggest that
the quark-gluon plasma may contain a developed network of magnetic flux tubes.
The existence of the vortex network may lead to observable effects in the
quark-gluon plasma because the chromomagnetic field of the vortices
should scatter and drag quarks.
\end{abstract}


\date{July 31, 2007}

\maketitle

\section{Introduction}
Studies of properties of thermal plasma became a major development in QCD in recent years, for a review see, e.g. \cite{plasmareview,latticereview}.
Properties of the plasma are studied both directly, at RHIC and via lattice simulations. On the theoretical side, novel ideas, like
AdS/CFT correspondence are being invoked \cite{gubser}, to say nothing of traditional approaches based
on various quasiparticle models~\cite{ref:phenomenology} and on field theory at finite temperature.

The traditional approach to the thermal plasma treats it, in zero approximation, as gas of free gluons and quarks and, then, takes
into account perturbative corrections.  An outcome of such calculations is a representation of the energy and pressure densities
as perturbative series in the effective coupling constant $g^{2}(T)$:
\begin{equation}
\epsilon(T)~= c_{\mathrm{SB}} \, T^{4} f_{\mathrm{pert}}(g^{2}(T))\,,
\label{perturbative}
\end{equation}
where $c_{\mathrm{SB}}$ is the Stefan-Boltzmann (SB) coefficient,
proportional to the number of degrees of freedom, and the perturbative expansion,
$f_{\mathrm{pert}}(g^{2}(T))$ is known explicitly up to terms of order $g^{6}\ln T$.
Perturbative predictions for such global characteristics of the plasma as energy density
turn to be in reasonable agreement with the data~\cite{latticereview}.

On the other hand, some particular properties of the plasma, such as
viscosity~\cite{viscosity}, indicate that, in the zero approximation,
plasma is to be considered rather as an ideal liquid than an ideal gas.
There is no yet a coherent picture that would unify both perturbative
and nonperturbative features of the QCD plasma.

It was speculated in Refs.~\cite{chernodub,chris,Shuryak:Liao} that there exists a
magnetic component of the Yang-Mills plasma which at temperatures not much above
the critical temperature $T_c$ is crucial for the plasma properties.
In Refs.~\cite{chris,Shuryak:Liao} constituents of the magnetic component
are thought to be classical magnetic monopoles. In Ref.~\cite{chernodub}
the magnetic component is identified with so-called magnetic strings related to magnetic monopoles.
The properties of the strings, or center vortices and their role in confinement have been
discussed in the lattice community for more than a decade, for review and
references see Ref.~\cite{greensite}.

According to the vortex picture the quark confinement emerges due to spatial percolation
of the magnetic vortex strings which lead to certain amount of disorder.
The value of the Wilson loop changes by a center element of the gauge group if the
magnetic vortex pierces the loop. Therefore, very large loops
receive fluctuating contributions from the vortex ensembles. These fluctuations make the average
value of the Wilson loop very small. One can show that the suppression of the loop follows
an area law for very large loops~\cite{greensite}.

Magnetic monopoles are certain gluonic objects that are
related to color confinement via the so-called dual superconductor
mechanism~\cite{DualSuperconductor}.
The condensate of the monopoles -- which is formed in the low-temperature phase --
expels the chromoelectric field exhibiting a dual analogue of the Meissner effect.
The chromoelectric field of quarks is squeezed into a dual
analogue of the Abrikosov vortex leading to quark confinement
(for a review see Ref.~\cite{ref:review:monopoles}).

\begin{figure}[htb]
\begin{center}
    \vskip 3mm \includegraphics[scale=0.6,clip=false]{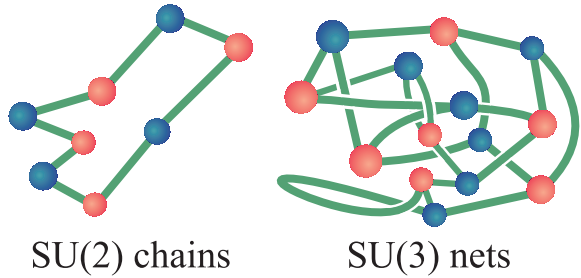}
\end{center}
\vskip -3mm \caption{(Color online) An illustration of the monopole-vortex chains and the monopole-vortex nets in
SU(2) and SU(3) Yang-Mills theories, respectively.}
\label{fig:nets}
\end{figure}

The presence of both the magnetic monopoles and the magnetic vortices in the vacuum of
Yang-Mills theory seem to be an indication of the existence of a more complicated genuine non-Abelian
object. Indeed, Abelian monopoles and center vortices appear to be strongly correlated
with each other~\cite{ref:chains}: almost all monopoles are sitting on top of vortices.
In $SU(2)$ gauge theory the genuine object is a monopole-vortex chain~\cite{ref:chains}. In
$SU(3)$ Yang-Mills theory one can expect existence of monopole-vortex 3-nets: the
junction of the three vortices may be viewed as a nexus~\cite{ref:net:analytical}
or/and as a center monopole~\cite{ref:net:numerical}. An illustration of the chains and nets
in a 3-dimensional timeslice is shown in Figure~\ref{fig:nets}.

The formation of these composite objects is essential for the
self--consistent treatment of monopoles in the quark-gluon plasma~\cite{chernodub}.
Similar monopole-vortex chains were found in numerous (non-)supersymmetric non-Abelian gauge theories
involving various Higgs fields~\cite{ref:monvort,ref:monvort:ab}. In particular, in Abelian gauge
models with a compact gauge field, monopoles and vortices should be organized in chains/nets
in order to be consistent with breaking of the confining electric string in specific representations
of the gauge group~\cite{ref:monvort:ab}. A short overview of the monopoles and vortices in
the Yang--Mills plasma can be found in Ref.~\cite{ref:review:MSU}.

In terms of the continuum theory, the
magnetic strings are defined as (infinitely thin) surfaces that are closed in
the vacuum state and can be open on an external 't Hooft line
(the same as the standard, or ``electric'' strings can be open on
the Wilson line). One can argue (for review and references see Ref.~\cite{vz}),
that the magnetic strings of the continuum theory can be identified
with the center vortices of lattice studies.  Moreover, it is also known that, indeed, properties of the vortices
are changing drastically once temperature is increased above
the critical value $T_c$ \cite{greensite,Langfeld:1998cz,zubkov,jeff:finite,langfeld2}.
In particular, they become time oriented
and this agrees with the assumption that the vortices become a component of the
Yang-Mills thermal plasma~\cite{chernodub}.

Once the magnetic component of the plasma is identified
with the magnetic vortices, further information on its properties
can be obtained by direct numerical calculations
on the lattice at finite temperature.
In this paper, we report on results of lattice calculations of contribution
of magnetic strings to the trace anomaly of energy-momentum tensor in the gluon plasma.
The trace anomaly is
intimately related to the equation of state, or energy density and pressure
of the thermal Yang-Mills plasma. Preliminary results of our investigation
were reported in Ref.~\cite{ref:Lattice}. Our lattice calculations refer to the
case of pure $SU(2)$ gauge theory which is conceptually similar to a more realistic
theory with three colors.

The structure of the paper is as follows. In Section~\ref{sec:YM:trace} we discuss basic
relations of Yang-Mills thermodynamics both in the continuum limit and on the lattice.
In Section~\ref{sec:moderate} we present our results for the thermodynamics
of $SU(2)$ Yang-Mills theory on moderately-sized lattice. Although the numerical
results on the $SU(2)$ thermodynamics are available in the literature~\cite{Engels:1988ph}
we repeat the calculations on lattices of particular geometries as these lattices are later used in
Section~\ref{sec:vortices} to calculate thermodynamical properties of magnetic vortices
such as the thermal vortex density and contribution to the (trace) anomaly of the energy
momentum tensor. A short summary of our results is given in Section~\ref{sec:summary} which
is followed by a concluding Section.

\section{Yang-Mills thermodynamics \newline and trace anomaly}
\label{sec:YM:trace}

\subsection{Continuum thermodynamics}

The free energy of an $SU(N)$ gauge system
\beqn
F = - T \log \cZ(T,V)\,,
\eeqn
is expressed via the partition function $\cZ$ as follows:
\beqn
\cZ = \int \dD A \, \exp\left\{- \frac{1}{2 g^2} \Tr \, G_{\mu\nu}^2 \right\}\,,
\label{eq:cZ}
\eeqn
where $G_{\mu\nu} = G_{\mu\nu}^a t^a$ is the field strength tensor of the non-Abelian field $A$, and
$t^a$ are the $SU(N)$ generators normalized in the standard way, $\Tr t^a t^b = \frac{1}{2} \delta^{ab}$.
The pressure $p$ is given by the derivative of the partition
function~\eq{eq:cZ} with respect to the spatial volume, $V$,
\beqn
p = \frac{T}{V} \frac{\partial \log Z(T,V)}{\partial \log V} = - \frac{F}{V} = \frac{T}{V} \log \cZ(T,V)\,.
\label{eq:pressure}
\eeqn
The last two equalities are valid for a sufficiently large and homogeneous system residing in
thermodynamical equilibrium.

The energy density $\epsilon$ is given by the derivative of the partition function~\eq{eq:cZ} with respect
to the temperature~$T$,
\beqn
\varepsilon = \frac{T}{V} \frac{\partial \log Z(T,V)}{\partial \log T}\,.
\label{eq:energy}
\eeqn
The relation between the pressure~\eq{eq:pressure} and the energy~\eq{eq:energy} constitutes
the equation of state of the system, $p = p(\varepsilon)$.

According to Eqs.~\eq{eq:pressure} and \eq{eq:energy} it is sufficient to determine the partition function of the
system in order to calculate the equation of state. However, available lattice simulations techniques are suitable
for the calculation of quantum averages of operators rather than for the evaluation of the partition function itself.
On the other hand, both the energy density $\varepsilon$ and the pressure $p$ can be derived from the quantum
average of a single quantity, which is the trace of the energy--momentum tensor~$T_{\mu\nu}$.

In $SU(N)$ gauge theory the energy--momentum tensor is given by the formula
\beqn
T_{\mu\nu} = 2 \Tr \left[G_{\mu\sigma} G_{\nu\sigma} - \frac{1}{4} \delta_{\mu\nu} G_{\sigma\rho} G_{\sigma\rho}\right]\,.
\label{eq:T}
\eeqn
At the classical level the energy-momentum~\eq{eq:T} is traceless because the {\it bare} Yang--Mills theory is a conformal theory.
However, at the quantum level the conformal invariance is broken, and, consequently, the energy--momentum tensor exhibits
a trace anomaly: the quantum average of the trace of the energy-momentum tensor,
\beqn
\theta(T) = \langle T^\mu_\mu \rangle \equiv \varepsilon - 3 p\,,
\label{eq:theta}
\eeqn
is nonzero. The trace anomaly is intimately related to the gluon condensate which breaks the scale invariance of the theory
\beqn
\theta(T) = \langle \frac{\beta(g)}{2 g} G_{\mu\nu}^a G_{\mu\nu}^a \rangle\,,
\label{eq:theta:continuum}
\eeqn
where the Gell-Mann-Low $\beta$-function is
\beqn
\beta(g) = \frac{\partial \, g}{\partial \log \mu} = - g^3 (b_0 + b_1 g^2 + \dots)\,,
\label{eq:beta:function}
\eeqn
with one- and two-loop perturbative coefficients
\beqn
b_0 = \frac{11 N}{3 (4\pi)^2}\,, \qquad b_1 = \frac{34 N^2}{3 (4\pi)^4}\,.
\label{eq:b0b1}
\eeqn

The thermodynamic relations in Eqs.~\eq{eq:pressure} and \eq{eq:energy}
relate the trace anomaly~\eq{eq:theta} with the pressure $p$,
\beqn
\theta(T) = T^5 \frac{\partial}{\partial T} \frac{p(T)}{T^4}
= - T^5 \frac{\partial}{\partial T} \frac{\log \cZ(T,V)}{T^3 V}\,.
\label{eq:anomaly:continuum}
\eeqn
Then, the pressure and, respectively, the energy density can be expressed via the trace anomaly, as follows:
\beqn
p(T) & = & T^4 \int\nolimits^T_0 \ \frac{\dd\, T_1}{T_1} \ \frac{\theta(T_1)}{T_1^4}\,,
\label{eq:pressure:anomaly}\\
\varepsilon(T) & = & 3 \, T^4 \int\nolimits^T_0 \ \frac{\dd\, T_1}{T_1} \ \frac{\theta(T_1)}{T_1^4} + \theta(T)\,.
\label{eq:energy:anomaly}
\eeqn
Equations \eq{eq:pressure:anomaly} and \eq{eq:energy:anomaly} demonstrate that the trace anomaly~\eq{eq:theta}
is the key quantity that allows us to reconstruct the whole equation of state.

The trace anomaly should vanish in a system of free massless relativistic particles
\beqn
\varepsilon^{\mathrm{SB}} = 3 p^{\mathrm{SB}} = N_{\mathrm{d.o.f.}} \frac{\pi^2}{30} T^4\,,
\label{eq:SB}
\eeqn
[where $N_{\mathrm{d.o.f.}} = 2 (N^2 -1)$ is the number of degrees of freedom in the noninteracting gas
of the $SU(N)$ gluons] or in the case when excitations are too massive compared with the temperature, $m \gg T$
(then $\varepsilon \sim p \sim \exp\{- m/T\}$). For Yang--Mills theory
these statements imply that the dimensionless quantity $\theta/T^4$ should approach zero at
very high temperatures (the gluons form a weakly interacting gas) and at very low temperatures
(note that in $SU(N)$ Yang--Mills theories the mass gap is much greater than the critical
temperature~\cite{Fingberg:1992ju}). The latter property is used to fix
the lower integration limits in Eqs.~\eq{eq:pressure:anomaly} and \eq{eq:energy:anomaly} at $T=0$.

\subsection{Lattice thermodynamics}

The lattice analogue of the partition function~\eq{eq:cZ} of $SU(N)$ gauge theory is
\beqn
\cZ(T,V) = \int D U \, \exp\Bigl\{ - \beta \sum_P S_P[U]\Bigr\}\,.
\label{eq:lattice:cZ}
\eeqn
The plaquette action $S_P[U]$ of the gluonic link fields $U_{x\mu}$ is usually written in the Wilson form,
\beqn
S_P[U] = 1 - \frac{1}{N} {\mathrm{Re}}\, \Tr U_P\,.
\label{eq:SP}
\eeqn
The identification of the lattice results with the physics in continuum
is achieved in the limit of vanishing lattice spacing, $a \to 0$:
\beqn
U_{x\mu} & = & \exp\Bigl\{i g \int\limits_x^{x+a \hat \mu} \dd y \, A_\mu(y) \Bigr\} \\
& \to & \bbbone + i a g A_\mu(x) + O(a^2)\,,
\nonumber
\eeqn
The lattice spacing $a$ is a function of the lattice coupling
\beqn
\beta = 2 N/g^2\,.
\eeqn

The spatial volume $V = (L_s a)^3$ and the temperature
\beqn
T = \frac{1}{L_t a}
\label{eq:temperature}
\eeqn
of the gluonic system are related to the asymmetric geometry of
the Euclidean lattice, $L_s^3 L_t$. The shorter direction, $L_t$ with $L_t \leqslant L_s$, is associated with the imaginary time,
or, ``temperature'' direction. The imaginary time formalism allows us to calculate various thermodynamic quantities
corresponding to the gauge system residing in thermodynamic equilibrium at given temperature and volume. Using the
relation $T (\partial/\partial T) = - a (\partial/\partial a)$ one can rewrite the continuum expression for the
anomaly~\eq{eq:anomaly:continuum} in a form suitable for numerical lattice simulations,
\beqn
\frac{\theta(T)}{T^4} = 6 \, L_t^4 \left(\frac{\partial \, \beta(a)}{\partial \log a} \right)
\cdot \Bigl(\langle S_P \rangle_T - \langle S_P \rangle_0\Bigr)\,.
\label{eq:anomaly:lattice}
\eeqn
Here the plaquette averages ${\langle S_P \rangle}_T$ and ${\langle S_P \rangle}_0$ are the action densities
taken, respectively, in a thermal bath at $T>0$ ($L_s^3 L_t$ lattices) and at $T=0$ ($L_s^4$ lattices).
In Eq.~\eq{eq:anomaly:lattice} it is implied that the $T=0$ plaquette expectation value is subtracted to
remove the effect of quantum fluctuations, which lead to an ultraviolet (UV) divergency of the quantum expectation
value. As a result, the trace anomaly becomes a UV-finite quantity, which is normalized
to zero at $T=0$ because of the existence of the mass gap.
Equation~\eq{eq:anomaly:lattice} is a lattice version of Eq.~\eq{eq:theta:continuum}.
The trace anomalies and the equation of state for $SU(2)$ and $SU(3)$ gauge theories were calculated
in Refs.~\cite{Engels:1988ph} and \cite{ref:Karsch:SU3}, respectively.

The anomaly~\eq{eq:anomaly:lattice} can be separated into electric and magnetic parts, respectively,
\beqn
\theta(T) = \theta_E(T) + \theta_M(T)\,.
\eeqn
The electric contribution comes from the field strength tensors of the chromoelectric fields,
$E^a_i = G^a_{i4}$, while the magnetic part is solely due to chromomagnetic fields,
$B^a_i = (1/2) \epsilon_{jk} G^a_{ijk}$. On the lattice the former is associated with temporal plaquettes,
$P_t \equiv P_{i4}$ with $i=1,2,3$, while the latter is related to the spatial
plaquettes, $P_s \equiv P_{ij}$ with $i<j=1,2,3$:
\beqn
\frac{\theta_E(T)}{T^4} & = & 3 \, L_t^4 \left(\frac{\partial \, \beta(a)}{\partial \log a} \right)
\cdot \Bigl(\langle S_{P_t} \rangle_T - \langle S_P \rangle_0\Bigr)\,,
\label{eq:anomaly:lattice:E}\\
\frac{\theta_M(T)}{T^4} & = & 3 \, L_t^4 \left(\frac{\partial \, \beta(a)}{\partial \log a} \right)
\cdot \Bigl(\langle S_{P_s} \rangle_T - \langle S_P \rangle_0\Bigr)\,.
\label{eq:anomaly:lattice:M}
\eeqn

Summarizing, the only lattice observable which is needed for the calculation of the trace anomaly~\eq{eq:anomaly:lattice}
[including its electric~\eq{eq:anomaly:lattice:E} and magnetic~\eq{eq:anomaly:lattice:E} parts] and, consequently, for
the determination of thermodynamics of Yang--Mills theory, is the difference between the
expectation values of the action densities calculated at zero-temperature (at $L_t = L_s$) and at finite
temperature (at $L_t < L_s$). The $\beta$--function in Eq.~\eq{eq:anomaly:lattice} has to be determined from
independent lattice simulations.

\section{SU(2) thermodynamics at moderate lattices}
\label{sec:moderate}
In this Section we present basic results for the thermodynamics of $SU(2)$ gauge fields. The
thermodynamics of $SU(2)$ lattice Yang-Mills theory has been studied in detail in Ref.~\cite{Engels:1988ph}.
In this Section we repeat certain numerical calculations of Ref.~\cite{Engels:1988ph} at moderate sized
lattices which will be used in the subsequent Section in the investigation of the vortex contributions to the
thermodynamics of the system.

We are working on the lattices $18^3\times 4$ and $18^4$ corresponding to finite- and
zero-temperature cases, respectively. At the asymmetric lattice with $L_t = 4$
the thermal phase transition is realized at $\beta = \beta_c \approx 2.3$,
Ref.~\cite{Engels:1989fz}. In physical units the transition in $SU(2)$ gauge theory is
achieved at temperature~\cite{Fingberg:1992ju} $T_c = 305(8)$~MeV, if one fixes the scale
by setting the $T=0$ tension of the chromoelectric string at the phenomenologically
accepted value of $\sigma^{1/2} = 440$~MeV.

The gauge field configurations are generated using the standard Wilson action~\eq{eq:SP}.
In order to determine the lattice spacing $a$ as a function of the lattice gauge coupling $\beta$
we use the interpolating function of Ref.~\cite{ref:beta:interpolation} which describes
with a high accuracy various lattice calculations of $a$. The lattice spacing $a$ in units of
the $T=0$ string tension $\sigma$ is parameterized as follows:
\beqn
\ln(a^2\sigma) = - \frac{64 \pi^4 b_0}{\beta}
+ \frac{b_1}{b_0^2} \ln\frac{64 \pi^4 b_0}{\beta}
+ \frac{d}{3 b_0 \beta} + c\,. \
\label{eq:inter}
\eeqn
The first two terms in the right hand side (r.h.s.) of the above equations represent,
respectively, the first and the second order loop corrections coming from the perturbation
theory. The coefficients $b_0$ and $b_1$ are the coefficients of the
$\beta$-function~\eq{eq:beta:function} which are given by Eq.~\eq{eq:b0b1} with $N=2$.
The other two terms in the r.h.s. of Eq.~\eq{eq:inter} are introduced to mimic higher-order
perturbative corrections as well as nonperturbative effects. It was
found~\cite{ref:beta:interpolation} that the choice
$c = 4.38(9)$ and $d = 1.66(4)$ reproduces measured valued of the lattice
spacing $a$ very well.

We use the interpolating relation~\eq{eq:inter} in order to determine the
temperature~\eq{eq:temperature} and the prefactor in the r.h.s. of
Eq.~\eq{eq:anomaly:lattice}. In Table~\ref{tbl:selected:beta} we present several
important quantities inherent to our numerical calculations.
\begin{table}
\begin{tabular}{|c|c|c|c|c|c|c|c|}
\hline
$\beta$      & 2.27  &  2.3   &   2.35  &  2.4   &   2.45 &   2.53  &  2.66    \\
\hline
$T/T_c$      & 0.909 &  1.005 &   1.188 &  1.401  &  1.651 &  2.14   &  3.24    \\
\hline
$a$, fm      & 0.183 &  0.165 &   0.140 &  0.119 &  0.101 &  0.0777 &  0.0513  \\
\hline
$L_s a$, fm  & 3.293 &  2.977 &   2.520 &  2.136 &  1.813 &  1.399  &  0.9234  \\
\hline
$ - \frac{\partial \beta}{\partial \log a}$
             & 0.2965 & 0.2982 & 0.3011 & 0.3038 & 0.3064 & 0.3103 & 0.3160 \\
\hline
\end{tabular}
\caption{Parameters of our numerical simulations: temperature (in units of the critical temperature~$T_c$),
lattice spacing~$a$, and spatial extension~$L_s a$ for selected values of the lattice coupling~$\beta$.}
\label{tbl:selected:beta}
\end{table}

\begin{figure}[htb]
\begin{center}
    \vskip 3mm \includegraphics[scale=0.25,clip=false]{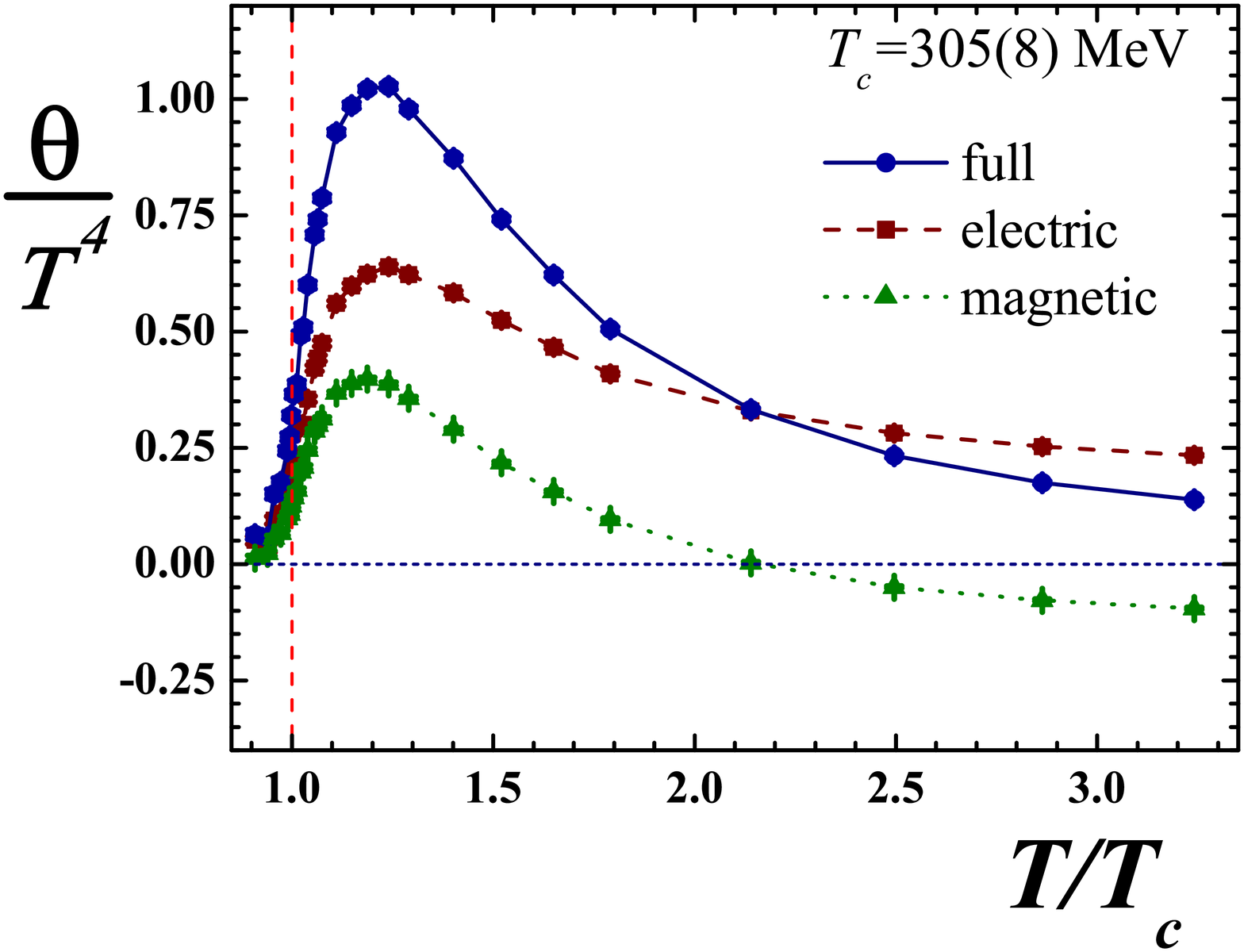}
\end{center}
\vskip -3mm \caption{(Color online) The trace anomaly $\theta$, represented as a dimensionless ratio $\theta/T^4$,
as a function of the temperature~$T$ (in units of the critical temperature, $T_c$).
We show the full anomaly (circles), and its electric (squares) and magnetic (triangles) contributions
determined by Eqs.~\eq{eq:anomaly:lattice}, \eq{eq:anomaly:lattice:E} and \eq{eq:anomaly:lattice:M},
respectively.}
\label{fig:su2}
\end{figure}

In Fig.~\ref{fig:su2} we show the trace anomaly~\eq{eq:anomaly:lattice} along with its electric~\eq{eq:anomaly:lattice:E}
and magnetic~\eq{eq:anomaly:lattice:M} parts.
It is very interesting to notice that the magnetic contribution to the anomaly vanishes in the deconfinement
phase at a temperature approximately twice larger than the transition temperature,
\beqn
\theta_M(T_0) = 0\,, \qquad T_0 =2.15(1) \, T_c\,.
\label{eq:T0}
\eeqn
This distinguished value of the temperature is also seen from the asymmetry of the so-called $A^2$--condensate:
the thermal fluctuations of the electric and magnetic components of the gluonic fields in the Landau gauge
becomes equal around $2 T_c$, see Ref.~\cite{ref:asymmetry} for details. In the case of the $SU(3)$ gauge group the
magnetic contribution to the trace anomaly also vanishes around the temperature which is slightly higher than
$2T_c$, Ref.~\cite{ref:Karsch:SU3}. The mentioned coincidence allows us to conjecture that the value~\eq{eq:T0}
of the critical temperature, $T \approx 2 T_c$, is in fact universal for all numbers of colors $N$.

\section{Thermodynamics of vortices}
\label{sec:vortices}

\subsection{Magnetic vortices and gluon trace anomaly: lattice definitions}

As we have discussed in the Introduction, (magnetic) center vortices are stringlike
configurations of the gauge fields which are relevant for color confinement in the
low-temperature confinement phase. In this Section we investigate how these objects contribute
to the thermodynamics of the $SU(2)$ gauge system.

The magnetic vortices carry a magnetic flux characterized by a nontrivial center element of
the gauge group. Such objects can conveniently be found in a specific gauge, which makes
the gauge field as close to the center subgroup of the gauge group as possible. In $SU(2)$ Yang-Mills theory
the relevant gauge -- often called as the (direct) maximal center gauge -- is defined by
a maximization of the functional
\beqn
\max_{\Omega} F[U^\Omega]\,, \qquad F[U] = \sum_l (\Tr U_l)^2\,,
\label{eq:F}
\eeqn
with respect to the gauge transformations
\beqn
U_{x\mu} \to U_{x\mu}^\Omega = \Omega_x U_{x\mu} \Omega^\dagger_{x+\hat\mu}\,.
\label{eq:omega}
\eeqn
In Eq.~\eq{eq:F} the sum goes over all links $l$ of the lattice.

The center element of the gauge field $U_l$ can be identified after
the $SU(2)$ maximal center gauge~\eq{eq:F} is fixed:
\beqn
Z_l = {\mathrm{sign}}\, \Tr U_l = \pm 1\,, \qquad Z_l \in \Z_2\,.
\label{eq:Zl}
\eeqn
The maximal center condition \eq{eq:F} fixes the gauge freedom~\eq{eq:omega} up to
a center subgroup of the $SU(2)$ gauge group:
\beqn
U_{x\mu} \to U_{x\mu}^\omega = \omega_x U_{x\mu} \omega_{x+\hat\mu}\,, \qquad \omega_x \in \Z_2\,.
\label{eq:Z2U}
\eeqn
In terms of the $\Z_2$ gauge variables~\eq{eq:Zl} the center transformation~\eq{eq:Z2U},
\beqn
Z_{x\mu} \to Z_{x\mu}^\omega = \omega_x Z_{x\mu} \omega_{x+\hat\mu}\,,
\label{eq:Z2:gauge}
\eeqn
can naturally be interpreted as a $\Z_2$ transformation in a $\Z_2$ effective gauge theory
written in terms of the gauge fields~\eq{eq:Zl}.

The field strength tensor of the center gauge fields~\eq{eq:Zl} is the $\Z_2$ plaquette
\beqn
Z_P = Z_1 Z_2 Z_3 Z_4\,, \qquad Z_P \in \Z_2\,,
\label{eq:ZP}
\eeqn
where the subscripts $1 \dots 4$ denote the links forming the border of the plaquette $P$.
The $\Z_2$--field strength tensor~\eq{eq:ZP} is invariant under the $\Z_2$ gauge
transformations~\eq{eq:Z2:gauge}.

The vortex worldsheets $\Sigma$ are identified with the help of the $\Z_2$-plaquette~\eq{eq:ZP}.
Let us introduce the notation ${}^*P$ for the plaquette which belongs to the dual lattice
and which is dual to the plaquette $P$. Then
the dual plaquette ${}^*P$ does not contain the center vortex if the corresponding plaquette
is center-trivial, $Z_P = +1$. Equivalently, one may say that no vortex
is going through the plaquette ${}^*P$, if $Z_P = +1$. However, if  $Z_P = -1$,
the plaquette ${}^*P$ is a part of the vortex world-sheet. In short,
\beqn
Z_P =
\left\{
\begin{array}{ll}
-1, & \mbox{if}\ P \in \Sigma \\
+1, & \mbox{if}\ P \notin \Sigma
\end{array}
\right.
\label{eq:ZP:Sigma}
\eeqn
One can easily prove that vortices are closed loops since their
worldsheets are closed surfaces.

Using Eq.\eq{eq:ZP:Sigma} one can define the lattice vortex density
\beqn
\rho_P = \frac{1}{2} (1 - Z_P) =
\left\{
\begin{array}{ll}
1, & \mbox{if}\ P \in \Sigma \\
0, & \mbox{if}\ P \notin \Sigma
\end{array}
\right.
\label{eq:ZP:rho}
\eeqn
Then the averaged lattice vortex density is
\beqn
\rho = \frac{1}{2} (1 - \langle Z_P \rangle)\,.
\label{eq:rho}
\eeqn

The lattice fields can be decomposed into the vortex singular contribution
and the rest
\beqn
U_l = Z_l \, {\tilde U}_l\,,
\label{eq:Ul:decomp}
\eeqn
where the $SU(2)$ gauge field $ {\tilde U}_l$ is defined as follows
\beqn
\Tr {\tilde U}_l >0\,.
\label{eq:Ul:decomp:cond}
\eeqn

The trace of the $SU(2)$ plaquette (which is an explicitly $SU(2)$ invariant quantity),
can be decomposed into a singular (center-valued) contribution and a regular contribution,
respectively,
\beqn
\Tr U_P = Z_P \, \Tr {\tilde U}_P\,, \qquad
{\tilde U}_P = {\tilde U}_1 {\tilde U}_2 {\tilde U}^\dagger_3 {\tilde U}^\dagger_4\,.
\label{eq:TrU:decomp}
\eeqn
This relation is very useful for us since it allows us to identify the contribution
of the singular magnetic vortices into the vacuum expectation value (vev) of
the $SU(2)$ plaquette $\langle \Tr U_P \rangle$. This quantity is related to the
vev of the gluonic action~\eq{eq:SP}, which, in turn, contributes to the
thermodynamics of the system via the trace anomaly~\eq{eq:anomaly:lattice}.

\begin{figure}[htb]
\begin{center}
    \vskip 3mm \includegraphics[scale=0.5,clip=true]{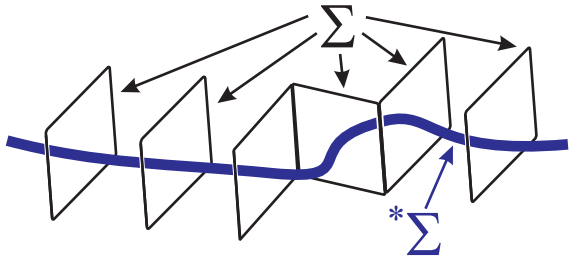}
\end{center}
\vskip -3mm \caption{(Color online) A three dimensional visualization of the magnetic vortex string~${}^*\Sigma$
which pierces a set of the plaquettes~$P \in \Sigma$.}
\label{fig:plaquettes}
\end{figure}

In order to calculate the contribution of magnetic
vortices into the trace anomaly~\eq{eq:anomaly:lattice}, and, consequently,
into the equation of state, we utilize the following chain of
considerations. For any given configuration of the gauge field $U_l$ one
can identify the location of the center vortex worldsheet, ${}^*\Sigma$,
as we described above. The worldsheet ${}^* \Sigma$ is a collection of
plaquettes ${}^* P$ of the dual lattice, which forms a closed surface, Figure~\ref{fig:plaquettes}.
Then the set of all plaquettes of the lattice can be separated into two
(generally, unequal) subsets: (i) the plaquettes $P$ of the original
lattice belonging to the vortex worldsheet, $P \in \Sigma$; and (ii) the plaquettes $P$ which
are outside of vortices, $P \notin \Sigma$.
The total action of the gauge field configuration can equivalently
be decomposed into the sum of the contributions of the individual plaquettes belonging
to the vortex worldsheet and the sum going over plaquettes outside of vortices, respectively:
\beqn
\sum_P S_P & = & \sum_{P \in \Sigma} S_P + \sum_{P \notin \Sigma} S_P \\
& \equiv & \sum_P \rho_P S_P + \sum_P (1 - \rho_P) S_P\,.
\label{eq:decomp}
\eeqn
Then we divide both sides of Eq.~\eq{eq:decomp} by the total number of
the lattice plaquettes, $6 L_s^3 L_t$, and average over all gauge field configurations.
As a result, we get that the average gluonic action per plaquette is given by the sum
\beqn
\langle S_P \rangle = \langle S_P \rangle^\vort + \langle S_P \rangle^\novort\,,
\label{eq:sum:SP}
\eeqn
where $\langle S_P \rangle^\vort$ is the vev of the gauge action coming from the vortex worldsheets,
\beqn
\langle S_P\rangle^\vort & = & \langle \rho_P S_P\rangle \equiv \frac{1}{6 L_s^3 L_t} \langle \sum_{P} \rho_P S_P \rangle
\label{eq:SP:vort}\\
& = & \rho + \frac{1}{4} \Bigl(\langle \Tr \widetilde U_P \rangle - \langle \Tr U_P \rangle \Bigr)\,,
\nonumber
\eeqn
and the action coming from the space unoccupied by the vortex worldsheets,
\beqn
\langle S_P\rangle^\novort & = & \langle (1 - \rho_P) S_P\rangle
\label{eq:SP:novort}\\
& \equiv  & \frac{1}{6 L_s^3 L_t} \langle \sum_{P} (1 - \rho_P) S_P \rangle
\nonumber\\
& = & 1 - \rho - \frac{1}{4} \Bigl(\langle \Tr \widetilde U_P \rangle + \langle \Tr U_P \rangle \Bigr)\,.
\nonumber
\eeqn

In order to derive the relations Eq.~\eq{eq:SP:vort} and \eq{eq:SP:novort} we used Eqs.~\eq{eq:ZP:rho}, \eq{eq:rho}
as well as Eq.~\eq{eq:SP} for the case of two colors, $N=2$. Then we utilized the relations,
\beqn
\rho_P S_P & = & \rho_P + \frac{1}{4} Z_P \Tr U_P - \frac{1}{4} \Tr U_P\,, \\
(1-\rho_P) S_P & = & (1-\rho_P) - \frac{1}{4} Z_P \Tr U_P - \frac{1}{4} \Tr U_P\,, \qquad
\eeqn
and used the fact that, according to Eq.~\eq{eq:TrU:decomp},
\beqn
Z_P \Tr U_P = \Tr \widetilde U_P\,.
\label{eq:tilde:UP}
\eeqn

It is important to realize that the vev of the quantity~\eq{eq:tilde:UP}
can be interpreted as the vev of the plaquette $\Tr U_P$ evaluated on
configurations with ``removed vortices''. Indeed, if we manually set $Z_P=1$ at all plaquettes
of the gauge field configuration, then the vacuum expectation of $\Tr U_P$ at these modified
configurations is automatically equivalent to the vev of $\Tr \widetilde U_P$ at
the original (unmodified) configuration:
\beqn
\langle \Tr \widetilde U_P \rangle \equiv \langle Z_P \Tr U_P \rangle \equiv \langle \Tr  U_P \rangle_{\mathrm{removed\ vortices}}
\label{eq:removed}
\eeqn
The procedure of ``removing vortices'' is equivalent to a shift of the gluonic link fields $U_l$,
\beqn
U_l \to Z_l \, U_l \equiv {\tilde U}_l\,, \qquad Z_l = {\mathrm{sign}}\, \Tr U_l\,,
\eeqn
according to the decomposition~\eq{eq:Ul:decomp}. By definition, the thus-modified configuration
contains no center vortices. Note that the gauge field configurations with removed
vortices are not able to support the color confinement~\cite{ref:removed:vortices}.

Summarizing, the lattice density of the gluonic action -- which is crucial for the determination
of the trace anomaly~\eq{eq:anomaly:lattice} -- can be split according to Eq.~\eq{eq:sum:SP}
into a piece coming from the vortex worldsheets~\eq{eq:SP:vort} and arising from the regions
of space lying outside of vortices~\eq{eq:SP:novort}. In the next Subsections we analyze these
contributions in more details.

\subsection{Center vortices and action density:\newline numerical results}

We have studied vortex properties on the lattices $18^4$ and $18^3\times 4$.
Depending on the value of the lattice coupling $\beta$ we used
from 100 to 500 statistically independent configurations for the zero
temperature lattice $18^4$, and from 700 to 1600 configuration
for the finite-temperature lattice $18^3\times 4$. Generally, the larger
$\beta$ the larger is the number of gauge field configurations needed to reach
an acceptable statistical accuracy of the numerical results. The maximal center
gauge~\eq{eq:F} was fixed using a simple iterative algorithm. In our
simulations we considered one Gribov copy only. Despite the unsophisticated
nature of our approach, we believe -- following the results reported in
Ref.~\cite{ref:center:vortex} and in Ref.~\cite{ref:cautionary} --
that this algorithm allows us to capture
not only qualitative but also, to a large extent, various quantitative features
of the vortex ensembles at finite temperature. In our visualization
of the vortex-related data we indicate the statistical errors only, leaving aside
the (systematical) error bars related to the gauge fixing issues such as the
choice of the Gribov copy.

Below we present results of our numerical simulations for several important quantities
calculated both in lattice and in physical units. Before proceeding to the calculation
of the trace anomaly and other related quantities, we discuss the important numerical
results for the density of the magnetic (center) vortices.

\subsubsection{Vortex density}
\label{sec:vortex:density}

At finite temperature the vortex density $\rho$ was studied in Ref.~\cite{Langfeld:1998cz,zubkov,jeff:finite,langfeld2}.
We calculate the vortex density both at finite temperature
and at zero-temperature in order to disentangle the finite-temperature effects from the
effects related to the variation of the UV cutoff. Indeed,
the UV cutoff -- the role of which is played by the (inverse) lattice spacing $a$ --
is controlled by the lattice coupling constant $\beta$, as, for example, illustrated
by the interpolating function~\eq{eq:inter}. The temperature is related to the extension of the
lattice in the short direction~\eq{eq:temperature}. The comparison of the lattice
results obtained at different lattice geometries and at the same fixed value of $\beta$ allows
us to single out effects of temperature variations on the vortex density.

\begin{figure}[htb]
\begin{center}
    \vskip 3mm \includegraphics[scale=0.27,clip=true]{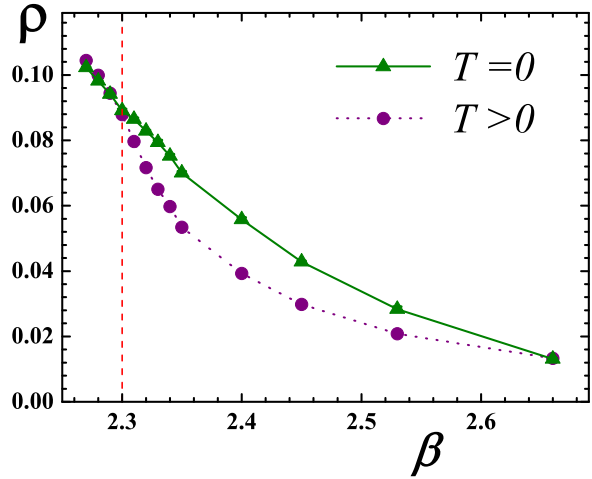}
\end{center}
\vskip -3mm \caption{(Color online) The lattice density of the center vortices~\eq{eq:rho} at zero-temperature (the solid line)
and at finite temperature (dotted line) vs the gauge coupling $\beta$. The simulations are done on $18^4$ and $18^3 \times 4$
lattices, respectively. The vertical dashed line marks the critical coupling
$\beta_c$ corresponding to the finite-temperature phase transition.
On the $18^3 \times 4$ lattice, a lower value of $\beta$ corresponds to a lower temperature.
Hereafter some sets of the numerical data points are connected by lines to guide the eye.}
\label{fig:density}
\end{figure}

In Figure~\ref{fig:density} we show the total density of the center vortices~\eq{eq:rho}
calculated on the lattices $18^4$ and $18^3\times 4$.
One can observe that in the confinement phase, $\beta < \beta_c \approx 2.3$,
the vortex density is almost the same for $L_t=18$ and $L_t=4$ lattices.

In the deconfinement phase, $\beta > \beta_c$ on  the $L_t = 4$ lattice, the vortex density starts
to deviate from its zero-temperature value (calculated on the $L_t=18$ lattice). It is quite surprising
to notice that just above the phase transition the density of the thermal vortices gets {\it lower}
compared with the value of the vortex density at zero-temperature. However, as
the temperature (or, equivalently, the value of the lattice gauge coupling $\beta$
on the $L_t = 4$ lattice) increases further, the absolute value of the difference in
the vortex densities at $L_t=4$ and $L_t=18$ lattices reaches its maximum, and then the deviation between
the vortex densities starts to decrease again. Thus, the effect of the temperature on the vortex
density is quite nontrivial.

At zero-temperature the vortex ensembles typically consist of one very long vortex worldsheet
which spreads over almost the whole spacetime, and a lot of short-sized clusters of vortex
worldsheets. The worldsheet of the infinitely long vortex trajectory is usually associated
with a condensed (infrared) vortex component while the short UV clusters form a
perturbative (UV) component. It is the condensate of vortices which leads to
the confinement of color charges. At $T > T_c$ the spatial string tension in the deconfinement phase
is supported by the vortex percolation in the spatial dimensions~\cite{Langfeld:1998cz,jeff:finite}.

At finite temperature it looks natural to separate the strings in thermodynamic ensembles into two components
describing, respectively, the virtual strings and the real strings. It is easy to explain the meaning of ``virtual''
and ``real'' in terms of particlelike objects which we understand much better than vortices which are stringlike
defects~\cite{chernodub}: the real objects are thermodynamically relevant while the virtual ones are not.
In other words, the virtual objects are associated with the ground state of the theory realized at zero-temperature while the
real particles are the excitations of this ground state.

Consider, for example, the Abelian monopoles at finite temperature
in Euclidean Yang-Mills theory. At zero-temperature a part of the
monopoles is condensed (similarly to vortices) while another part of
the monopole trajectories belong to ultraviolet clusters. At finite temperature the ensembles of
the monopole trajectories should contain both the virtual monopoles belonging to the vacuum and, at the same
time, the thermodynamically relevant (or, real) monopoles. In order to measure physically meaningful observables
in the continuum limit, one should be able to separate these types of monopoles from each other.
According to Ref.~\cite{chernodub} the thermodynamically relevant monopoles can be distinguished
from the virtual monopoles by a simple principle: the thermal monopoles should wrap around the
temperature (compactified) direction of the Euclidean space. Moreover, the quantum density of
the thermal monopoles is not equal to the average density of the total length of the (wrapped)
monopole trajectories as one could na\"ively guess on general grounds. Such a quantity is
divergent in the UV regime making its interpretation somewhat obscure. The density of
the thermal monopoles corresponds to the density of the winding number $s$ of the monopole trajectories
in the temperature direction of the Euclidean spacetime:
\beqn
\rho^{\mathrm{thermal}}_{\mathrm{mon}} = \frac{1}{V_{3d}} \langle |s| \rangle\,.
\label{exactt}
\eeqn

Coming back to the magnetic vortices, one can ask the important question: how to separate the thermal component of the vortex density from the
density of the virtual component? In other words: in Euclidean lattice simulations at finite temperature
we observe a set of closed vortex trajectories in each configuration of the gauge fields.
Which vortex worldsheets are real (thermal) and which worldsheets corresponds just to virtual strings? Following our experience with
particlelike monopoles~\cite{chernodub} one can suggest that the worldsheets of the real vortices are characterized by a nontrivial
wrapping number with respect to the (compactified) temperature dimensions.

A vortex analogue of Eq.~\eq{exactt} remains to
be derived. Note that the naive suggestion for the thermal vortex density -- the averaged area of the wrapped vortex worldsheets per
unit Euclidean volume -- in general should not be correct. In fact, in Ref.~\cite{chernodub} it was shown that the
analogous quantity for the monopoles (the average length of wrapped monopole trajectories per unit Euclidean volume) is --
contrary to the correct expression~\eq{exactt} -- divergent in the UV limit and thus cannot serve as the density of the
real monopoles.

\begin{figure}[htb]
\begin{center}
    \vskip 3mm \includegraphics[scale=0.27,clip=true]{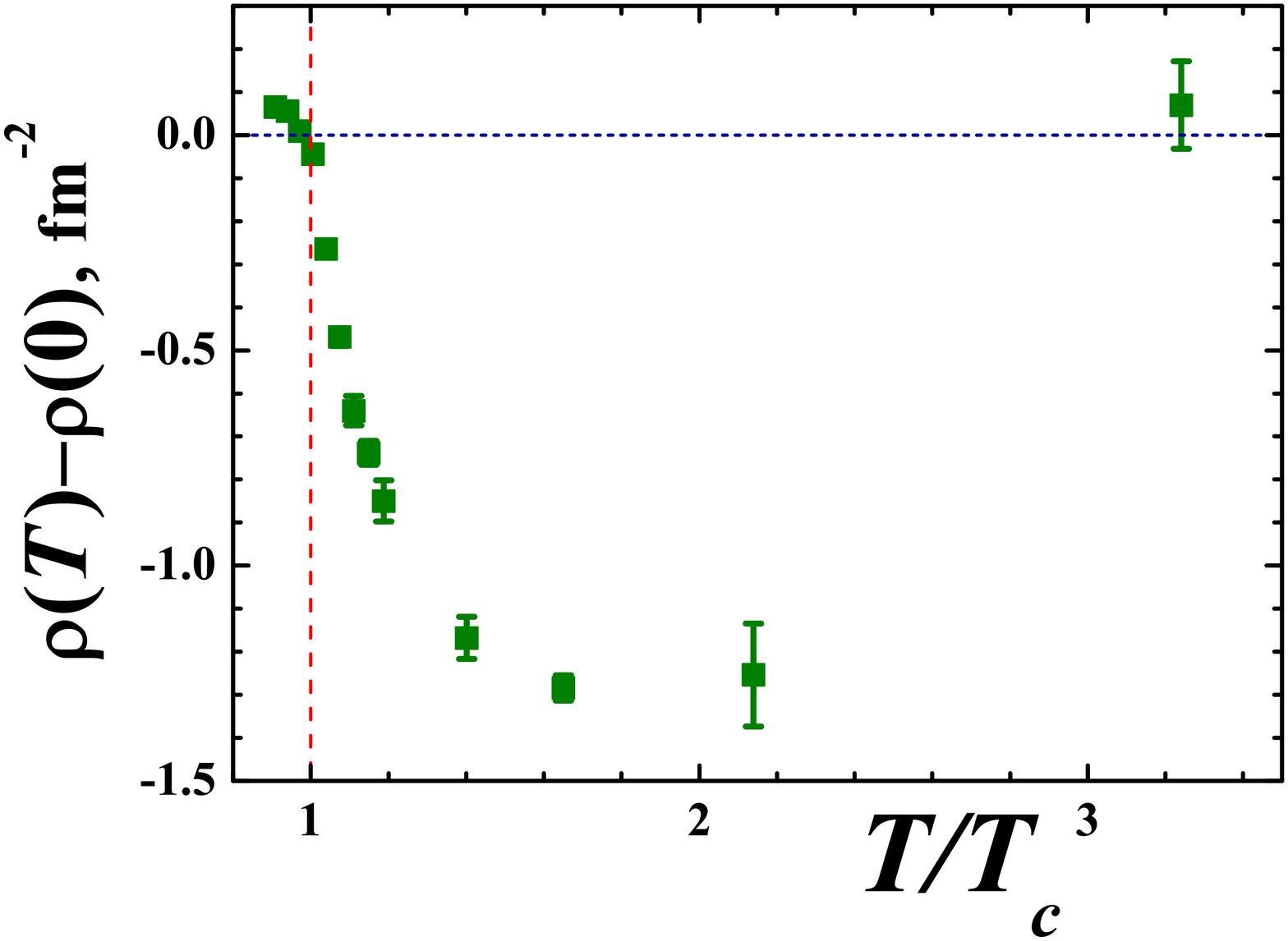}
\end{center}
\vskip -3mm \caption{(Color online) The difference between the vortex density at finite temperature and at zero-temperature (physical units).}
\label{fig:density:diff}
\end{figure}

Let us stress again a somewhat puzzling behavior of the vortex density, Figure~\ref{fig:density}.
Naively, in a quantum field theory an expectation value of a local quantity at finite temperature
should receive contributions from both quantum field fluctuations -- which are inherent solely to the
zero-temperature ground state -- and from various thermal excitations. The (virtual) quantum field fluctuations
have the same structure both at zero and at finite temperature (for example, the UV divergencies both
at zero and at finite temperatures are identical~\cite{ref:Kapusta}). Therefore, in order to figure out
a contribution of the thermal excitations one can fondly subtract a zero-temperature vev from the
thermal expectation value. This subtraction, for example, is routinely performed for the trace anomaly of
the Yang-Mills fields~\eq{eq:anomaly:lattice}. The puzzle is that if the subtraction procedure is applied to the
vortex density, then the would-be thermal vortex density takes a {\it negative} value in the deconfinement phase,
Figure~\ref{fig:density:diff}. The negative valued quantity is difficult to interpret in terms of the density of realistic objects.
Thus, the density of the thermal (real) vortices needs to be defined using a nonstandard way like it is done for the
particlelike objects~\eq{exactt} in Ref.~\cite{chernodub}. Note, that in our simulations we do not discriminate the
thermal (wrapped) and the UV vortex components.

One can observe a few interesting features of the difference in the vortex densities shown in Figure~\ref{fig:density:diff}:
\begin{itemize}
\item The difference in the vortex densities is positive in the confinement phase, $T < T_c$.
\item This quantity is negative in the deconfinement phase up to temperatures
of the order of, approximately, three critical temperatures, $T_c < T \lesssim  3 T_c$.
At higher temperatures the quantity will again take positive values.
\item The minimum of the thermal density is reached at $T \approx 2T_c$. Note that this is the temperature
at which the contribution of the magnetic gluons vanishes, see Eq.~\eq{eq:T0} and Figure~\ref{fig:su2}.
\end{itemize}

\subsubsection{Comment on monopoles vs vortices}

\begin{figure}[htb]
\begin{center}
    \vskip 3mm \includegraphics[scale=0.54,clip=true]{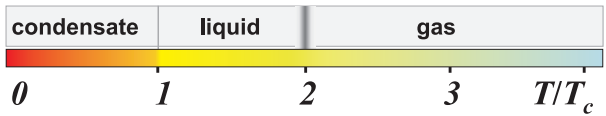}
\end{center}
\vskip -3mm \caption{(Color online) The states of the monopole matter in Yang-Mills theory at finite temperature according to Ref.~\cite{chernodub}.}
\label{fig:states}
\end{figure}

The fact that the minimum of the vortex density is reached at $T \approx 2T_c$ is intriguing. As we have discussed
above, magnetic vortices are related to magnetic monopoles. According to Ref.~\cite{chernodub} the
condensed (in the confinement phase) state of the monopoles melts into a monopole liquid at $T=T_c$ which is
evaporated into a monopole gas at $T \approx 2 T_c$, at which the vortex density reaches its minimum. The
coincidence of the temperatures may be not accidental. An illustration of the monopole states
in the finite-temperature Yang--Mills theory is provided in Figure~\ref{fig:states}.

The existence of the monopole liquid close to the critical point in the deconfinement phase
was confirmed by accurate lattice data of Ref.~\cite{ref:Italians} according to an analysis presented in
Ref.~\cite{Shuryak}. The appearance of the monopole gas at even higher temperatures -- discussed
earlier in analytical terms in Ref.~\cite{chris} -- was demonstrated numerically in Ref.~\cite{ref:blocking}.
We refer the interested reader to Ref.~\cite{ref:review:MSU} for a short review of this subject.

At very high temperatures the general behavior of the vortex density can be predicted
from dimensional reduction arguments. According to the dimensional reduction,
at high temperatures nonperturbative physics is controlled by three-dimensional magnetodynamics which corresponds
to zero Matsubara frequency of the original four-dimensional theory. The vortex trajectories
become static, and the vortex density becomes equal to the density of the vortex lines in the $3d$ space.
The magnetodynamical quantities are expressed in terms of the corresponding powers of the gauge coupling
in the dimensionally reduced theory,
\beqn
g^2_{3d}(T) = T \, g^2_{4d}(T)\,,
\eeqn
where $g_{4d}$ is the running coupling of the original four-dimensional theory calculated at the scale $T$.

Thus, density of vortices in the high-$T$ limit should be described by the formula,
\beqn
\rho(T) = C_\vort \, g^4_{3d}(T) \propto {\left(\frac{T}{\log T/\Lambda}\right)}^2\,, \qquad T \gg T_c\,, \
\label{density2}
\eeqn
where $C_\vort$ is a temperature-independent parameter and $\Lambda \sim \Lambda_{\mathrm{QCD}}$ is a dimensional parameter.

The power, with which the temperature enters Eq.~\eq{density2}, is clear from the dimensional arguments, $\rho \sim O(T^2)$.
However, the logarithmic behavior, $\log^{-\alpha_\vort}(T/\Lambda)$, with $\alpha_\vort = 2$ is governed by the perturbation theory.
It is interesting to note that the same power of the logarithm, $\alpha \approx 2$,
was recently found for the Abelian monopoles in the high-temperature phase of the Yang--Mills theory~\cite{ref:Italians}.
This fact is amusing, because for the monopoles, which are particlelike objects, the natural power of the logarithm
would be $\alpha_{\mathrm{mon}} = 3$, Ref.~\cite{chernodub,chris}. The observed difference
allowed the authors of Ref.~\cite{ref:Italians} to conclude that the monopoles do not form a free particle gas. We support this
opinion by noticing, that the monopoles are to be related to the magnetic vortices, and the vortex dynamics make the monopole
properties less trivial. This could explain why the observed power of the monopole logarithm, coincides with the vortex one,
\beqn
\alpha_{\mathrm{mon}} \approx \alpha_\vort = 2\,.
\eeqn

\subsubsection{Expectation value of SU(2) plaquettes}

We have calculated numerically the expectation values of the $SU(2)$ plaquettes $\frac{1}{2}\Tr U_P$ at the
original configurations and compared them with the vortex-removed configurations. The comparison was done
both at zero and finite temperature. The plaquette expectation values are crucial for the determination of
contribution of vortices into the expectation value of the gluon action, Eqs. \eq{eq:SP:vort},
\eq{eq:SP:novort}, and \eq{eq:removed}. Notice, that according to Eq.~\eq{eq:SP}, the larger the plaquette expectation
value the smaller the action density (and vice versa).

\begin{figure}[htb]
\begin{center}
    \vskip 3mm \includegraphics[scale=0.27,clip=true]{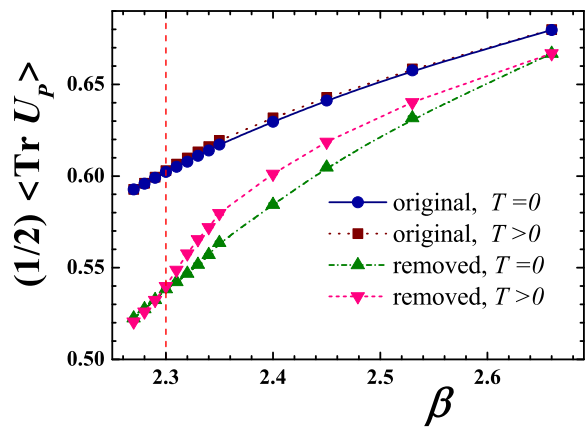}
\end{center}
\vskip -3mm \caption{(Color online) The expectation values of the plaquette $\frac{1}{2}\Tr U_P$ at zero and at finite temperatures
for usual and for vortex-removed configurations.}
\label{fig:utu}
\end{figure}

The expectation value of the $SU(2)$ plaquettes is presented in Figure~\ref{fig:utu}.
One can immediately observe the following interesting features:\\[3mm]
$\bullet$ The plaquette expectation values for configuration with removed vortices is much smaller compared
to the original unmodified configurations. This feature is valid both at zero and at finite temperatures in
all phases.
\begin{itemize}
\item[-] Thus, magnetic vortices carry positively valued energy density.
This fact was already established for $T=0$ Yang--Mills theory in Ref.~\cite{ref:fine:tuning}.
\end{itemize}
$\bullet$ The temperature fluctuations increase the plaquette expectation value both at the original configurations
and at the modified gauge field configurations.
\begin{itemize}
\item[-] Thus, the effect of the temperature is to decrease the energy density both at
the vortex worldsheets and at the space outside of vortices.
\end{itemize}
$\bullet$ The influence of the temperature on the plaquette expectation values of the vortex-removed configurations
is much stronger compared with the observed influence of the temperature on the original gauge field configurations.
\begin{itemize}
\item[-] Thus, the energy densities at the vortex worldsheets and outside of
vortices receive very strong contributions due to thermal configurations. These contributions are canceled in
the original configurations while they cannot be canceled in the modified configurations due to the removal of the vortices.
\end{itemize}
This observation stresses the significance of the vortices for the thermodynamics of the gauge system.

\begin{figure}[htb]
\begin{center}
    \vskip 3mm \includegraphics[scale=0.27,clip=true]{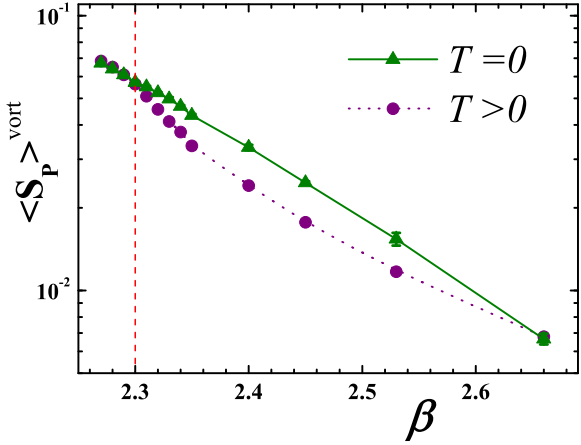}
\end{center}
\vskip -3mm \caption{(Color online) The expectation value of the action density calculated at the vortex worldsheets~\eq{eq:SP:vort} at zero
temperature ($18^4$ lattice, the solid line) and at finite temperature ($18^3 \times 4$ lattice, the dotted line)
vs the gauge coupling $\beta$.}
\label{fig:action:vort}
\end{figure}

Using the numerical data for the vortex density, for the plaquette expectation values both at the original configurations
and at the vortex-removed configurations one can calculate the contribution of the vortices to the vacuum expectation
value of the gluon action density. The contribution of the vortices -- calculated with the help of Eq.~\eq{eq:SP:vort} -- is
presented in Figure~\ref{fig:action:vort}. It is clearly seen that the effect of the temperature is to decrease the total
action carried by the vortices.

It is very important to stress, that we have observed two different temperature effects which are realized {\it just above}
the critical temperature in the deconfinement phase ($T_c < T \lesssim 2 T_c$):
\begin{itemize}
\item According to Figure~\ref{fig:action:vort} the higher the temperature the lower the contribution of the vortices to the
vev of the action density.
\item According to Figure~\ref{fig:density} the higher the temperature the lower the vortex density.
\end{itemize}

The knowledge of these facts alone does not allow us to conclude what is the primary reason of the diminishing of the vortex contribution into
the total action density. For example, the energy density per unit area of the vortex worldsheet
could be insensitive to the temperature and the overall effect could be due to the lowering of the vortex density with
increase of temperature. Other options are: the energy density at the worldsheet may moderately increase (or, decrease)
with temperature weakening (enhancing) the effect of the dropping vortex density.
The same question may be addressed at higher temperatures, $T \gtrsim 2 T_c$, at which both the vortex density
and the thermal plaquette values are increasing functions of temperature.

We study numerically
the action density per an elementary plaquette belonging to the vortex worldsheet\footnote{For the sake of
completeness one may suggest to perform even more refined measurement by separating the contributions from
the ultraviolet and the infrared vortex clusters following Ref.~\cite{ref:fine:tuning}. In our approach we
have not done this refinement treating both the UV and IR vortices on the same footing.}:
\beqn
\langle s_P\rangle^\vort = \frac{\langle S_P\rangle^\vort}{\rho} \equiv
\frac{\langle \sum_{P} \rho_P S_P \rangle}{\langle \sum_{P} \rho_P \rangle}\,,
\label{eq:s:vort}
\eeqn
the action density per an elementary plaquette outside of the vortex worldsheet
\beqn
\langle s_P\rangle^\novort = \frac{\langle S_P\rangle^\novort}{1 - \rho} \equiv
\frac{\langle \sum_{P} (1 - \rho_P) S_P \rangle}{\langle \sum_{P} (1 - \rho_P) \rangle}\,.
\label{eq:s:novort}
\eeqn
For the sake of brevity we call the quantity~\eq{eq:s:vort} as ``the specific vortex action'' because
it represents the gluonic action density normalized by the area of the vortex worldsheet. Analogously,
we refer to the quantity~\eq{eq:s:vort} as to ``the specific no-vortex action''.

The results -- which are represented in Figure~\ref{fig:action:density} - are quite interesting:
\begin{figure}[htb]
\begin{center}
    \vskip 3mm \includegraphics[scale=0.27,clip=true]{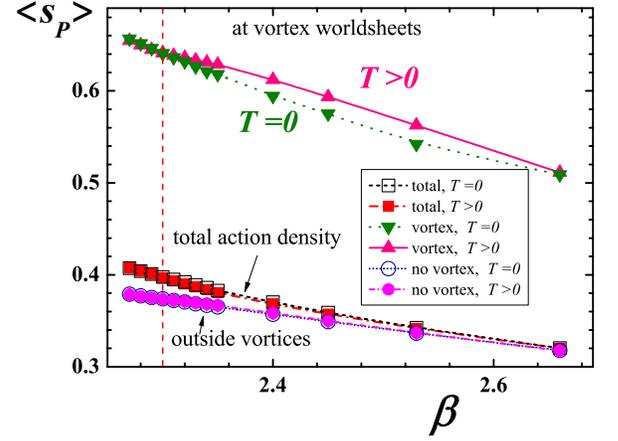}
\end{center}
\vskip -3mm \caption{(Color online) The same as in Figure~\ref{fig:action:vort} but for the action density
per an elementary plaquette belonging to the vortex worldsheet~\eq{eq:s:vort} and per an elementary plaquette
lying outside of vortex worldsheets~\eq{eq:s:novort}.}
\label{fig:action:density}
\end{figure}
\begin{itemize}
\item In the low temperature region of the deconfinement phase
($T_c < T \lesssim 3 T_c$) the specific vortex action is larger compared with
its value at $T=0$.
\item In all phases the specific vortex action is larger compared with the total action and, consequently, to the
specific no-vortex action.
\item The specific vortex action is very sensitive to the temperature variations compared with other action densities.
\end{itemize}

The last observation again indicates that the vortices play a very important role in the thermodynamics of
Yang-Mills theory. We discuss this issue in the next Subsection.

\subsection{Center vortices and trace anomaly:\newline  numerical results}

The lattice calculation of the trace anomaly requires a careful renormalization.
The renormalization of the total gluonic anomaly is quite straightforward~\eq{eq:anomaly:lattice}:
in order to cancel zero-point oscillations the expectation value of the $SU(2)$ plaquette calculated
at zero-temperature should be subtracted from the finite-temperature plaquette action. According to
a common wisdom, both these quantities contain the same UV divergencies, and the subtraction
leaves us with the UV-finite thermal contribution\footnote{Note, however, the cautionary remark
in Section~\ref{sec:vortex:density} about the renormalization of the vortex density: the naively
``renormalized'' vortex density turns out to be negative, Figure~\ref{fig:density:diff}.}.

The regularization of the vortex contribution is less straightforward. According to Eq.~\eq{eq:sum:SP}
the gluonic action can be split into two pieces corresponding to the vortex contribution, $\langle S_P \rangle^\vort$
and to the rest, $\langle S_P \rangle^\novort$, both at zero and at finite temperature. However, the quantum
fluctuations of the gluonic fields at the vortex and outside of vortex should not, in general, be the same because
the vortices are infinitely thin objects possessing very strong gluonic fields~\cite{ref:fine:tuning}.
Taking into account that the vortex density is a lively function of temperature,
Figure~\ref{fig:density:diff}, we conclude that the UV divergencies of, say, the action density $\langle S_P \rangle^\vort$
at $T=0$ and $T>0$ should not, in general, match. Therefore, the renormalized finite-temperature
value of $\langle S_P \rangle^\novort$ by a subtraction of the same quantity at $T=0$ should give, in general,
a UV-divergent result.

In order to illustrate a possible failure of a naive renormalization, we take the lattice expression for the trace
anomaly~\eq{eq:anomaly:lattice}, and separate the vortex contribution from the rest as follows:
\beqn
\langle S_P \rangle_T - \langle S_P \rangle_0 & = &
\left[\langle S_P \rangle^\vort_T - \rho(T) \langle S_P \rangle_0\right]
\nonumber\\
& & \hskip -15mm
+ \left\{\langle S_P \rangle^\novort_T - \left[1-\rho(T)\right] \langle S_P \rangle_0\right\}\,.
\label{eq:split:SP}
\eeqn
Here we used Eq.~\eq{eq:sum:SP} which implies that the vev of the action can exactly be split into the vortex contribution,
$\langle S_P \rangle^\vort$, and the contribution outside of vortices, $\langle S_P \rangle^\novort$. This renormalization
subtracts from the vortex-originated $\langle S_P \rangle^\vort_T$ the appropriate amount of the $T=0$
gauge action, $\langle S_P \rangle_0$. The coefficient of proportionality is given by the vortex density
$\rho$: each vortex plaquette at $T>0$ is regularized by the plaquette action at $T=0$. The same type of the
regularization is done for the no-vortex contribution, $\langle S_P \rangle^\novort$.

According to Eqs.~\eq{eq:anomaly:lattice} and \eq{eq:split:SP} the trace anomaly can then be split into
the two regularized contributions
\beqn
\theta(T) = \theta^{\vort}_{\mathrm{naive}}(T) + \theta^{\novort}_{\mathrm{naive}}(T)\,,
\eeqn
where
\beqn
\frac{\theta^{\vort}_{\mathrm{naive}}(T)}{T^4} & = & 6 \, L_t^4 \left(\frac{\partial \, \beta(a)}{\partial \log a} \right)
\nonumber\\
& & \cdot \left[\langle S_P \rangle^\vort_T - \rho(T) \langle S_P \rangle_0\right]\,,
\label{eq:vort:naive} \\
\frac{\theta^{\novort}_{\mathrm{naive}}(T)}{T^4} & = & 6 \, L_t^4 \left(\frac{\partial \, \beta(a)}{\partial \log a} \right)
\nonumber\\
& & \cdot \left\{\langle S_P \rangle^\novort_T - \left[1-\rho(T)\right] \langle S_P \rangle_0\right\}\,.
\label{eq:novort:naive}
\eeqn

\begin{figure}[htb]
\begin{center}
    \vskip 3mm \includegraphics[scale=0.27,clip=true]{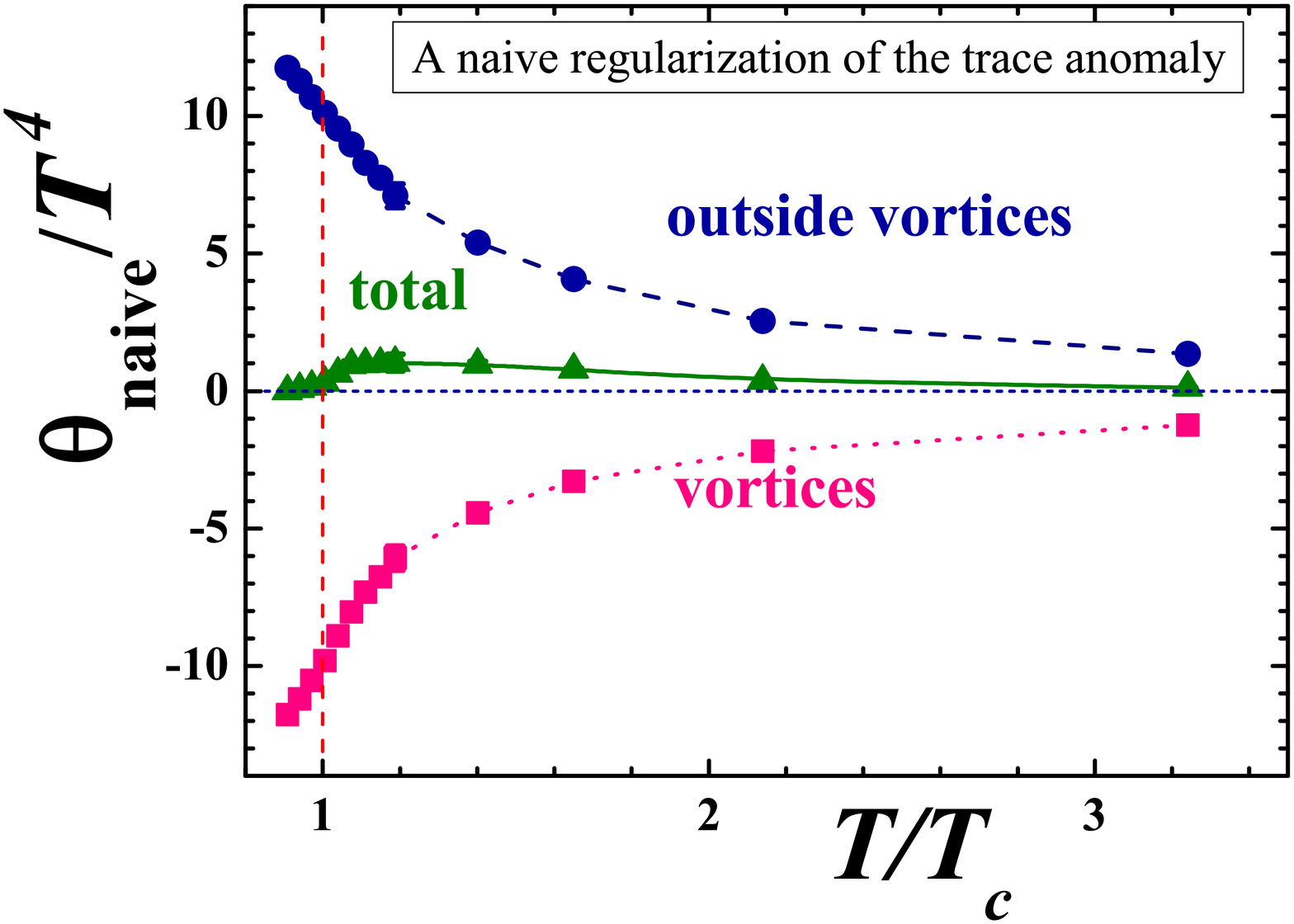}
\end{center}
\vskip -3mm \caption{(Color online) Naively regularized contribution of vortices into the trace anomaly~\eq{eq:vort:naive} vs temperature
(the data points are connected by the dotted line).
The contribution to the anomaly coming from the space outside of vortices~\eq{eq:novort:naive}, and the total
gluonic trace anomaly~\eq{eq:anomaly:lattice} are also shown (the dashed line and the solid line, respectively).
}
\label{fig:anomaly:naive}
\end{figure}

The results for the naively regularized quantities \eq{eq:vort:naive} and \eq{eq:novort:naive} are presented
in Figure~\ref{fig:anomaly:naive} (yet another simple regularization was implemented in our earlier
investigation in Ref.~\cite{ref:Lattice}). We see from Figure~\ref{fig:anomaly:naive} that the vortices
provide a large negative contribution while the
space outside of vortices is characterized by a large positive-valued contribution. The sum of the two gives
the total trace anomaly. The naively regularized vortex and no-vortex results are not sensitive to the phase
transition and they continue to rise as the temperature is lowered. Thus, we conclude the regularization
schemes \eq{eq:vort:naive} and \eq{eq:novort:naive} are not appropriate.

It seems that the right quantity which characterizes the thermal contribution of the vortices to the action density is
the specific vortex action~\eq{eq:s:vort}, i.e. the vortex action per unit area of the vortex worldsheet. The
regularization (i.e., the  subtraction of the zero-point fluctuations at zero-temperature) should be done accordingly:
the action per vortex plaquette at $T>0$ should be regularized by the action per vortex plaquette at $T=0$:
\beqn
\langle s_P\rangle^\vort_{\mathrm{reg}} = \langle s_P\rangle_T - \langle s_P\rangle_0\,,
\label{eq:s:vort:reg}
\eeqn
where the specific vortex action $\langle s_P\rangle$ is defined in Eq.~\eq{eq:s:vort}.

One can define the specific contribution $\vartheta^\vort$ of the vortices into the trace anomaly, or, the contribution of the magnetic vortices
to the trace anomaly counted per unit area of the vortex world surface at given temperature. Formally, the definition of this quantity is
\beqn
\vartheta^\vort = \frac{{\mathrm{d}}\,\theta^\vort}{{\mathrm{d}}\,{\mathrm{Area}}}\,,
\label{eq:vort:spec:formal}
\eeqn
where ``Area'' denotes the area of the vortex worldsheet at given temperature. The definition of this quantity
can be written using Eqs.~\eq{eq:anomaly:lattice} and \eq{eq:s:vort:reg}:
\beqn
\frac{\vartheta^\vort(T)}{T^4} = 6 \, L_t^4 \left(\frac{\partial \, \beta(a)}{\partial \log a} \right)
\cdot \Bigl(\langle s_P\rangle_T - \langle s_P\rangle_0 \Bigr)\,.
\label{eq:vort:specific}
\eeqn

\begin{figure}[htb]
\begin{center}
    \vskip 3mm \includegraphics[scale=0.27,clip=true]{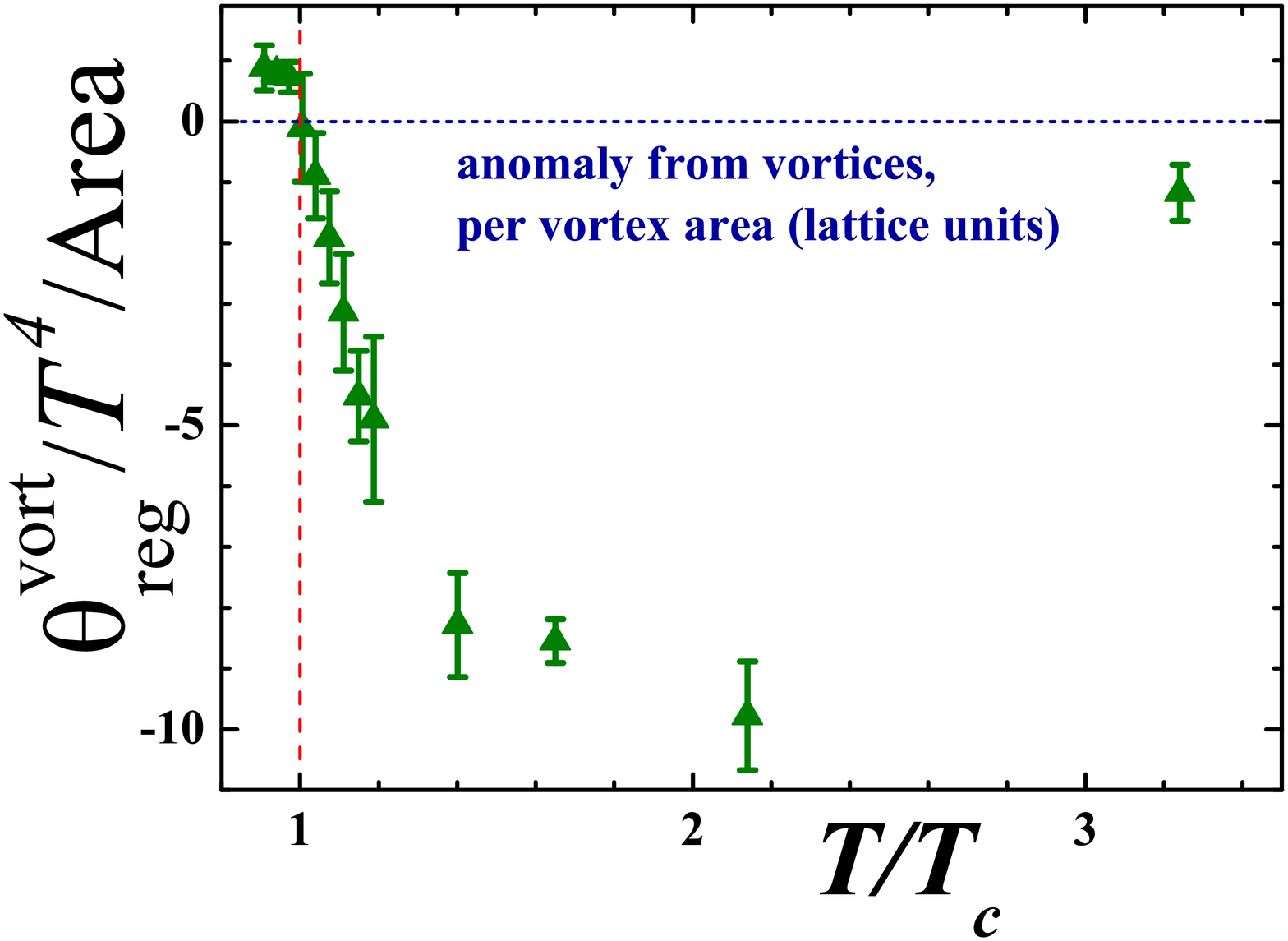}
\end{center}
\vskip -3mm \caption{(Color online) The specific anomaly, Eqs.~\eq{eq:vort:specific} and \eq{eq:vort:spec:formal}, coming from the magnetic vortices (lattice units).}
\label{fig:anomaly:vort:lat}
\end{figure}

We show the specific contribution of the magnetic vortices to the trace anomaly in Figure~\ref{fig:anomaly:vort:lat}. There are a few
interesting features of this quantity: in the deconfinement phase $T_c < T \lesssim 3.5 T_c$ the specific anomaly is negative. In the
deeper deconfinement phase as well as in the confinement phase the specific trace anomaly is positive. At the phase transition the specific
anomaly is close to zero. All these features are in qualitative agreement with our preliminary calculations reported in Ref.~\cite{ref:Lattice}.

The negative (ghostlike) value of the vortex-originated anomaly is not unexpected according to the theoretical
analysis of Ref.~\cite{ref:nonAbelian:vortices}. The vortices can be treated as nonabelian strings with a nontrivial
worldsheet dynamics associated, in particular, with the monopoles, localized on worldsheets of the magnetic vortices.

\begin{figure}[htb]
\begin{center}
    \vskip 3mm \includegraphics[scale=0.27,clip=true]{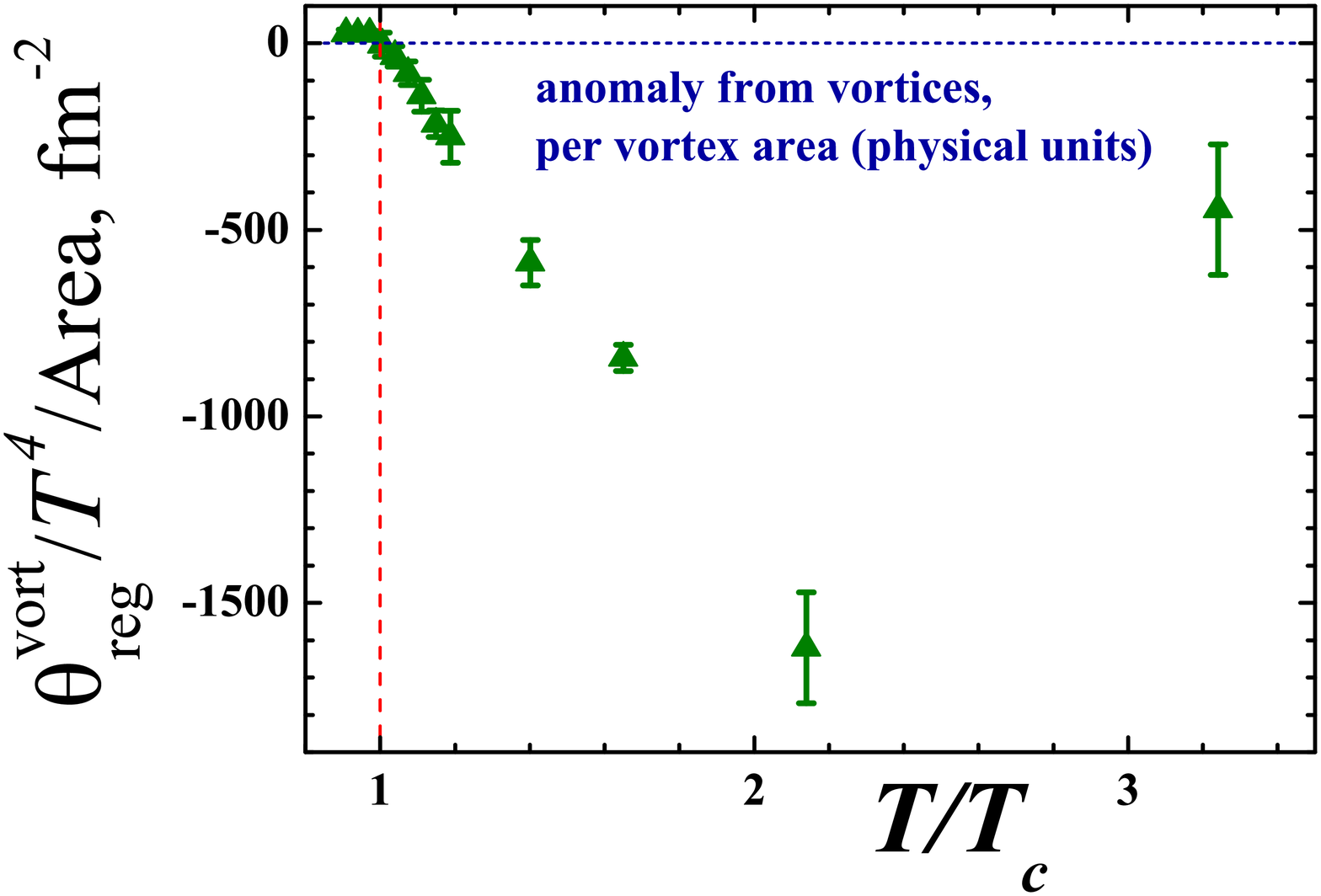}
\end{center}
\vskip -3mm \caption{(Color online) The same as in Fig.~\ref{fig:anomaly:vort:lat} but in physical units.}
\label{fig:anomaly:vort:phys}
\end{figure}

In physical units the specific vortex contribution to the anomaly is represented in Figure~\ref{fig:anomaly:vort:phys}.
The vortex contribution is numerically very large. This fact agrees qualitatively with a singular
nature of the distribution of the action density around vortex worldsheets~\cite{ref:fine:tuning}.

\begin{figure}[htb]
\begin{center}
    \vskip 3mm \includegraphics[scale=0.27,clip=true]{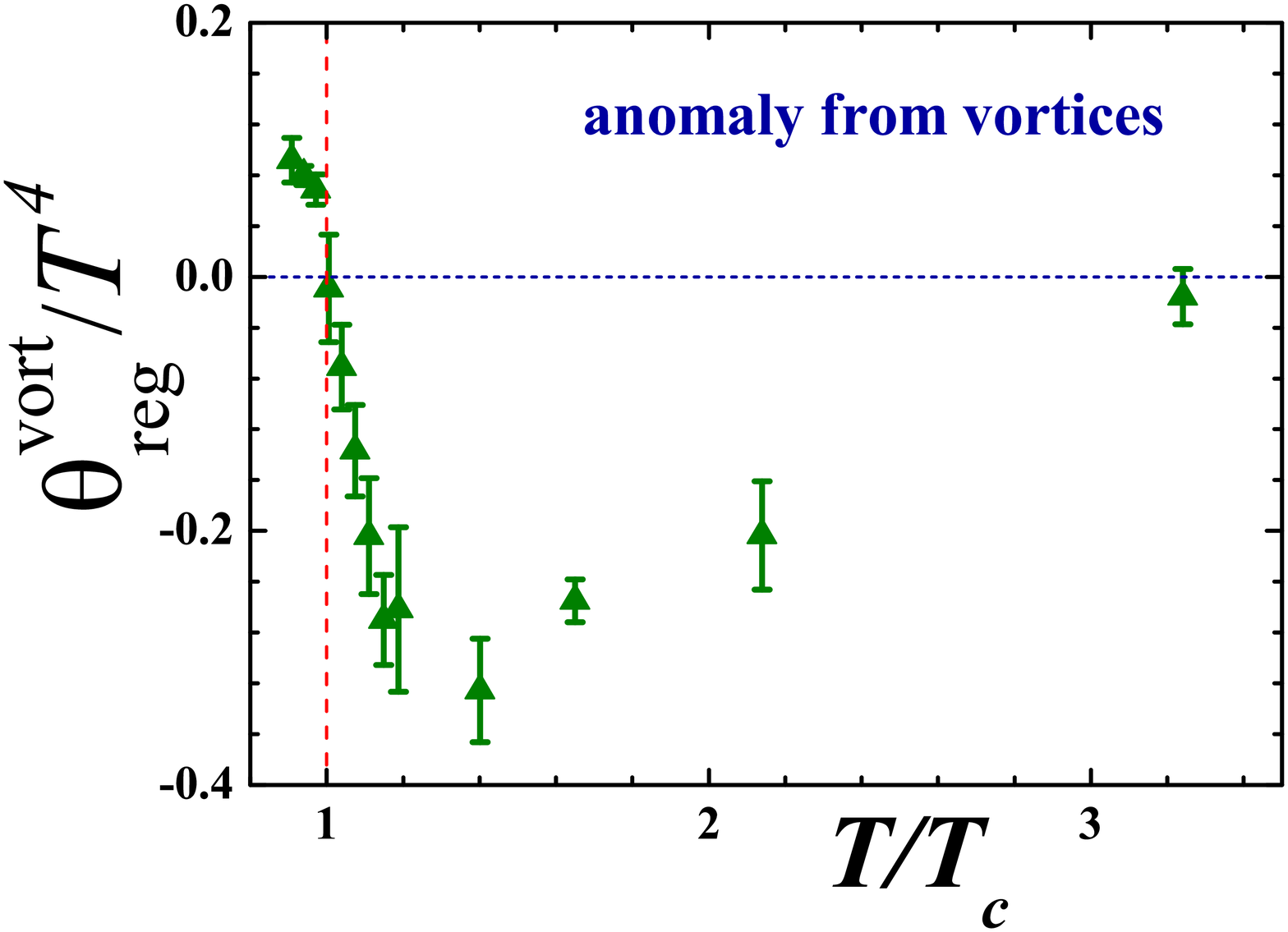}
\end{center}
\vskip -3mm \caption{(Color online) The contribution of the magnetic vortices into the trace anomaly, given by Eqs.~\eq{theta:vort} and \eq{eq:vort:specific}.}
\label{fig:anomaly:vort:total}
\end{figure}

Finally, the total regularized contribution of the vortices into the trace anomaly,
\beqn
\theta^\vort(T) = \rho(T) \, \vartheta^\vort(T)\,,
\label{theta:vort}
\eeqn
is represented in Figure~\ref{fig:anomaly:vort:total}.
The scale of the vortex-originated contribution
is useful to compare with the total gluonic anomaly, Figure~\ref{fig:su2}.
Some features of the total vortex contribution to the anomaly (Figure~\ref{fig:anomaly:vort:total}) are very similar to the specific
trace anomaly coming from vortices (Figure~\ref{fig:anomaly:vort:phys}): in the confinement phase, $T< T_c$, the vortex trace
anomaly is a positive quantity while in the deconfinement phase this quantity becomes negative in the region of
temperatures $T_c < T \lesssim 3 T_c$. As for higher temperatures,  $T \gtrsim 3 T_c$, our data does not allow us to discriminate
between the two possibilities: the contribution to the trace anomaly coming from the vortices may be either positive or negative, and the
ratio $\theta^\vort/T^4$ in this region may be very small. Note that due to the definition
of the vortex-originated anomaly, Eqs.~\eq{theta:vort} and \eq{eq:vort:specific}, this quantity should be equal to zero at $T=0$.
At the phase transition the anomaly is close to zero too.

It is appropriate to notice that at finite temperature some vortex trajectories may wrap with respect to the temporal (temperature)
direction while other vortices may have zero wrapping index. Following Ref.~\cite{chernodub} we suggest here that the wrapped vortices
have a direct relation to the thermodynamics of the system while the unwrapped vortices bear features of the zero-temperature theory only.
Thus we expect that the vortices with different wrappings with respect to the temporal direction may give different contributions
to the gluonic anomaly. In our simulations we do not discriminate between the vortices with different wrapping properties.

Certain qualitative features of the vortex anomaly, Figure~\ref{fig:anomaly:vort:total}, may in principle be related to certain geometrical
(orientational) properties of the vortex worldsheets. Indeed, it is well known~\cite{Langfeld:1998cz,jeff:finite}, that in the
deep confinement phase, $T \ll T_c$, the worldsheets of the magnetic vortices have no preferred orientations in the lattice spacetime
because there is practically no difference between spatial and temporal (temperature) directions. However, as the temperature increases the
preferences do appear: the asymmetry ratio between the densities of the spacelike (${}^* P_{ij}$) and timelike (${}^* P_{i4}$, $i,j=1,2,3$)
vortex worldsheets slowly rises in the confinement phase up to a critical point~\cite{Langfeld:1998cz,jeff:finite}.
In the deconfinement phase the ratio falls down drastically since
at $T \gg T_c$ the vortices are getting static according to the dimensional reduction arguments. On the other hand, we know that the
chromomagnetic (spacelike, $\theta_M \sim \Tr U_{P_{ij}}$) part of the anomaly is always smaller than the chromoelectric (timelike,
$\theta_E \sim \Tr U_{P_{i4}}$) part, Figure~\ref{fig:su2}. Since the spacelike
vortices (${}^* P_{ij}$) pierce timelike ($P_{i4}$) plaquettes and vise versa, the slow rise of the spacelike portion of the vortex
worldsheets in the confinement phase leads to an increase of the r\^ole of the timelike plaquettes in the vortex-dominated anomaly.

The timelike (chromoelectric) part of the {\it bulk} anomaly is bigger than the chromomagnetic part according to Figure~\ref{fig:su2}.
{\it If} the same is true for the anomaly at the vortex worldsheet then the gradual
orientation of the magnetic vortices towards spatial directions with the rise of temperature should lead to an increase of the vortex-originated
anomaly in accordance with our observations (Figure~\ref{fig:anomaly:vort:total}). As the system passes the deconfining point, the timelike
vortices become more favored compared with the spacelike ones and therefore the spacelike chromomagnetic plaquettes become dominant in
the vortex-originated contribution. Thus, taking into account the relation $\theta_M < \theta_E$ and using the geometrical arguments,
one can qualitatively explain a drop of the vortex-originated anomaly just above the critical temperature, Figure~\ref{fig:anomaly:vort:total}.

It is also important to stress that just above the phase transition the trace anomaly coming
from the vortices (Figure~\ref{fig:anomaly:vort:total}) is approximately of the same order as
the total gluon anomaly (Figure~\ref{fig:su2}). This fact highlights the significance of the
vortex degrees of freedom to the thermodynamical behavior of the gluon plasma.

Our numerical calculations do not allow us to trace the behavior of the vortex-originated trace anomaly in the confinement phase
at low enough temperatures, $T < 0.8 T_c$. As this quantity is still not zero at low temperatures, we cannot calculate the
contribution of the vortices into the pressure~\eq{eq:pressure:anomaly} and into the energy density~\eq{eq:energy:anomaly} because
they involve integration over the low temperature region. More refined and accurate calculations are needed to address this issue properly.

\begin{figure}[htb]
\begin{center}
    \vskip 3mm \includegraphics[scale=0.45,clip=true]{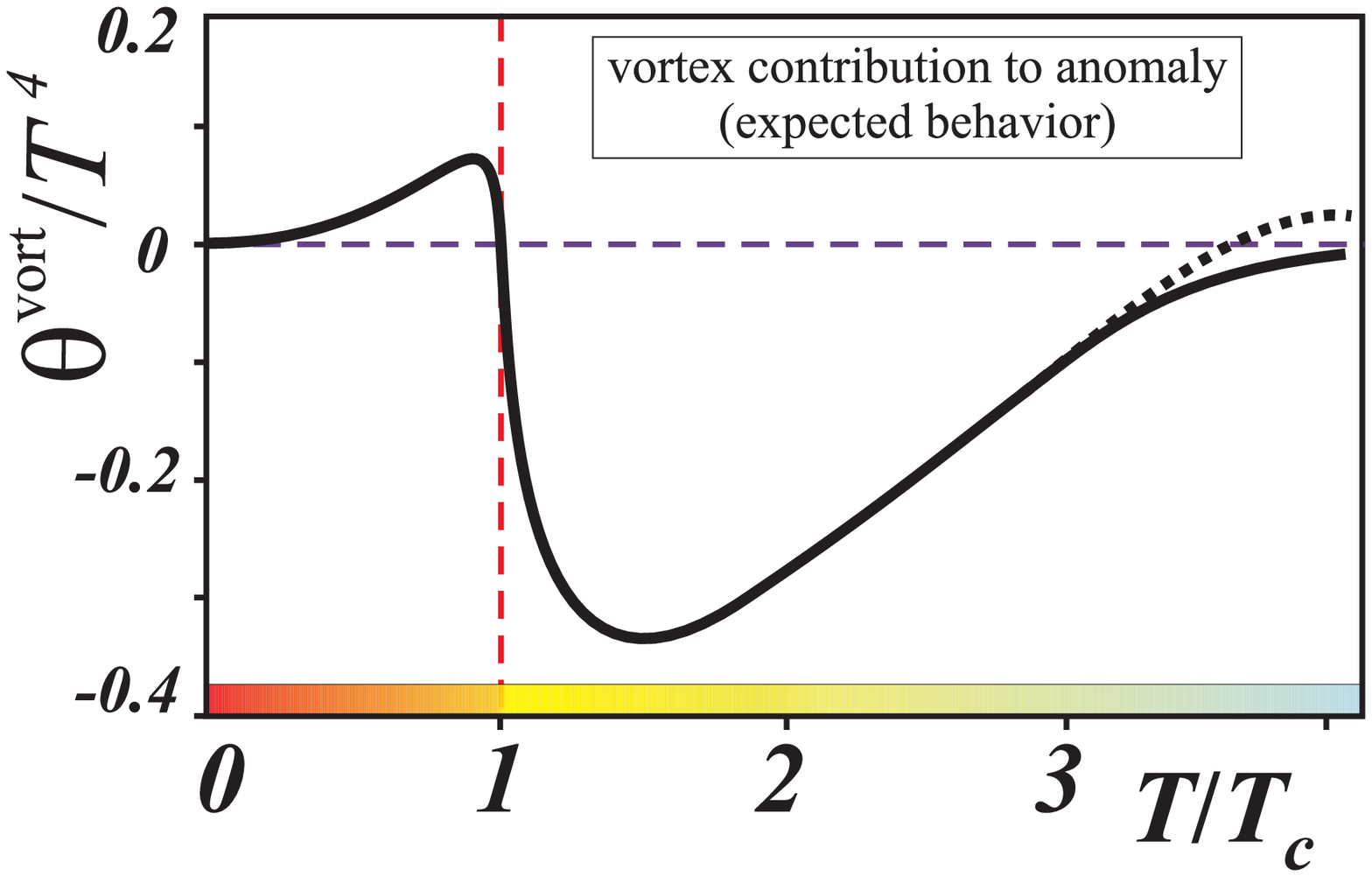}
\end{center}
\vskip -3mm \caption{(Color online) An expected behavior of the contribution of the magnetic vortices into the gluonic trace anomaly (the solid line).
The dotted line represents another high-temperature behavior that cannot be excluded by the available data.}
\label{fig:anomaly:vort:suggested}
\end{figure}

A qualitative extrapolation of our data into the low-temperature domain of the phase diagram allows us to suggest the
behavior of the vortex contribution to the trace anomaly as it is visualized in Figure~\ref{fig:anomaly:vort:suggested}.

\section{Summary of numerical results}
\label{sec:summary}

We observe the intriguing relevance of magnetic vortices to the thermodynamics of Yang--Mills theory.
A brief summary of the vortex properties is as follows:

\subsection*{Vortex density, Figure~\ref{fig:density:diff}}
\begin{itemize}
\item $T < T_c$. In the cold region of the confinement phase $\rho(T)$ is an increasing function of temperature~$T$.
As the temperature gets higher, the increase of the density turns into a sharp fall.
\item $T = T_c$. At the phase transition the vortex density is close to its zero
temperature value, $\rho(T_c) \approx \rho(0)$.
\item $T_c < T \lesssim 2 T_c$. Right above the critical temperature the vortex density continues
to fall and the minimum of the vortex density is reached approximately around $T \approx 2 T_c$. At this
temperature the magnetic contribution to the trace anomaly vanishes, Figure~\ref{fig:su2}.
\item $2 T_c \lesssim T \lesssim 3 T_c$.
The vortex density start to rise, and at $T \approx 3 T_c$ the density reaches its
zero-temperature value again, $\rho(3 T_c) \approx \rho(0)$.
\item $T \gtrsim 3 T_c$. In this domain of the temperatures the density is a positively defined rising quantity.
\item $T \gg T_c$. At very high temperatures the vortex density should follow the law of Eq.~\eq{density2}.
\end{itemize}
Our results on the vortex properties are in a qualitative agreement with other studies of the vortex properties
at finite temperature~\cite{Langfeld:1998cz,zubkov,jeff:finite,langfeld2}.
The rise and fall of the vortex density with the increase of temperature in the confinement phase should be understood
as a result of the competition of two opposite processes: the thermal fluctuations (i) excite the vortex media leading
to creation of additional virtual vortex loops from the vacuum; (ii) destroy the chaotic percolation of the vortex
cluster in the temporal direction making the vortex trajectories more static.
The percolation of the vortex trajectories in the temporal direction is not a well defined
quantity because of the finiteness of the extent of the Euclidean spacetime in this direction. Nevertheless, at low
temperatures not all the vortex loops wind around this direction and the increase of the temperature tries to break the
loops into pieces.

\subsection*{Vortex contribution into the trace anomaly, Figures~\ref{fig:anomaly:vort:total} and \ref{fig:anomaly:vort:suggested}}
We found that vortices do contribute to the equation of state of the gluon plasma.
The vortex contribution is illustrated in Figures~\ref{fig:anomaly:vort:total} and \ref{fig:anomaly:vort:suggested}.
Qualitatively, the temperature behavior of the vortex trace anomaly is very similar to the behavior of the
vortex density described above. We observe that:
\begin{itemize}
\item in the confinement phase the vortex trace anomaly is positively defined,
\item at the critical temperature, $T=T_c$, the vortex-originated trace anomaly is consistent with zero,
\item in the low-temperature domain of the deconfinement phase ($T_c < T \lesssim 3.5 T_c$) the trace anomaly coming from the vortices is negative,
\item at the high-temperature region of the deconfinement phase ($T \gtrsim 3.5 T_c$) the contribution of the vortices to the
gluon trace anomaly may either become positive or stay negative. Anyway, the ratio $\theta^\vort(T)/T^4$ is expected to be small in this region,
\item qualitatively, the features of the specific vortex trace anomaly (Figures~\ref{fig:anomaly:vort:lat}
and \ref{fig:anomaly:vort:phys}) and the ``bulk'' vortex trace anomaly (Figure~\ref{fig:anomaly:vort:total})
are very similar to each other,
\item the gluon action density at the vortex trajectories (Figure~\ref{fig:action:density}) is much higher compared with the gluon
action density outside of vortex worldsheets. This result is an agreement with previous studies reported in Ref.~\cite{ref:fine:tuning}, and
\item the value of the gluon action density at the vortices is much more sensitive to the temperature variations compared with the
gluon action density outside of vortices (Figure~\ref{fig:action:density}).
\end{itemize}

One can characterize the very strong contribution of the vortices into the trace anomaly and, consequently,
to the equation of the state of the Yang--Mills plasma as follows:
on our lattices the vortices occupy just 2{-}4\% of the lattice spacetime (Figure~\ref{fig:density}) in the deconfinement
phase while their regularized contribution to the trace anomaly (Figure~\ref{fig:anomaly:vort:total}) is of the order of the
total contribution of all gluons in the bulk of the system (Figure~\ref{fig:su2}).

\section{Conclusions}

The basic idea of this paper is to calculate the contribution of vortices to the vev of the Euclidean
action density of the gluonic fields. The action density, which is basically the gluon condensate, is related to
the thermodynamic quantities of the Yang-Mills theory -- such as pressure and energy density -- via the anomaly of
the trace of the energy--momentum tensor. A short summary of the results is presented in the previous Section.

We observed that magnetic vortices play an essential r\^ole in the thermodynamics of Yang-Mills theory.
Our main result is presented in Figure~\ref{fig:anomaly:vort:total}. In the temperature range
$T_c < T \lesssim 3 T_c$ the contribution
of the magnetic vortex strings to the trace anomaly is numerically large and {\it negative} in sign compared
to the total energy and the trace of the energy-momentum tensor of the whole of the plasma:
\begin{equation}
\theta_\vort(T) < 0\,, \quad \langle \Tr G^{2}(T)\rangle_\vort >  0\,, \quad T_c < T \lesssim 3 T_c \,.
\quad
\label{ghost}
\end{equation}
Taken at face value, Eqs.(\ref{ghost}) amount to the observation of a {\it ghostlike} component of the gluonic plasma.

The existence of magnetic vortices may strongly influence the physics of quarks in the
quark-gluon plasma. The random chromomagnetic fluxes of the string networks in the gluon plasma
may trap, scatter and drag the quarks. The chromoelectric and chromomagnetic flux tubes are suggested to
arise in the early Glasma regime of the evolution of color gauge fields in high energy heavy ion
collisions~\cite{ref:Larry}. We leave to a future publication~\cite{ref:preparation}
discussions of the magnetic string effects in thermally excited gluon media.

There are, however, reservations to be made. First, the theoretical interpretation of
data referring to a component of the plasma is subject to an uncertainty
because the magnetic component obviously interacts strongly with
the rest of the plasma. And it is rather the total energy that has a
direct physical meaning than the distribution of it. Thus, a detailed interpretation of the
data (\ref{ghost}) asks for further efforts on the continuum-theory side.
Second, it might be worth emphasizing that,
observationally, the negative sign is due to a decrease of the
density of the strings (compared with the $T<T_c$ case).
At this moment, it is not clear whether this mechanism, behind the signs
in Eq.(\ref{ghost}) is important for  the interpretation.

In any case, the numerical calculations reveal that magnetic strings, which at
our lattices occupy (2-4)\% of the total volume are crucial
for the dynamics of the whole plasma (in the limit of the vanishing lattice spacing $a\to 0$ this fraction
tends to zero). The absolute value of their contribution
to the energy of the plasma is of the order of the total value of the energy in the bulk of the system.
Thus, perturbative calculations, like (\ref{perturbative}), agree with the data averaged over the whole plasma
while nonperturbative dynamics is responsible for a highly nontrivial
sharing of the same energy density between various components of the plasma.
A remote analogy is a quark-resonance duality. While simple quark
graphs  describe well the cross section averaged over
a large energy interval, resonances are responsible for
a more local-in-energy structure.
It is becoming obvious that a detailed understanding of the plasma properties is
not possible without clarifying the role of its magnetic
component~\cite{chernodub}.

\begin{acknowledgments}
This work was supported by Grants-in-Aid for Scientific Research from "The Ministry
of Education, Culture, Sports, Science and Technology of Japan" Nos. 17340080 and 20340055,
by the STINT Institutional grant IG2004-2 025,
by the grants RFBR 06-02-04010-NNIO-a, RFBR 08-02-00661-a, DFG-RFBR 436 RUS,
by a grant for scientific schools NSh-679.2008.2, by the Federal Program of the
Russian Ministry of Industry, Science and Technology No. 40.052.1.1.1112
and by the Russian Federal Agency for Nuclear Power.
The numerical simulations were performed using a SX-8 supercomputer at
RCNP at Osaka University, and a SR11000 machine at Hiroshima University.
M.N.Ch is thankful to the members of the Department of Theoretical Physics of Uppsala University
for kind hospitality and stimulating environment. M.N.Ch and V.I.Z appreciate their
fruitful stays at the Institute for Theoretical Physics of Kanazawa University and the Research
Institute for Information Science and Education of Hiroshima University, Japan.
\end{acknowledgments}


\begin{thebibliography}{99}

\bibitem{plasmareview}
E.~V.~Shuryak, in ``Minneapolis 2006, Continuous advances in QCD'', p.~3,
Ed. by M. Peloso and M. Shifman (Singapore, World Scientific, 2007)
[hep-ph/0608177];
%%CITATION = HEP-PH/0608177;%%
J.~P.~Blaizot,
%``Theoretical overview: Towards understanding the quark-gluon plasma,''
J.\ Phys.\ G {\bf 34}, S243 (2007) [hep-ph/0703150].
%%CITATION = JPHGB,G34,S243;%%

\bibitem{latticereview}
F.~Karsch,
%``Properties of the quark-gluon plasma: A lattice perspective,''
Nucl.\ Phys.\  A {\bf 783}, 13 (2007)
[hep-ph/0610024];
%%CITATION = NUPHA,A783,13;%%
  U.~M.~Heller,
  %``Recent progress in finite temperature lattice QCD,''
  PoS {\bf LAT2006}, 011 (2006)
  [hep-lat/0610114].
  %%CITATION = POSCI,LAT2006,011;%%

\bibitem{gubser}
J.~J.~Friess, S.~S.~Gubser, G.~Michalogiorgakis, S.~S.~Pufu,
  %``Expanding plasmas and quasinormal modes of anti-de Sitter black holes,''
  JHEP {\bf 0704}, 080 (2007)
  [hep-th/0611005].
  %%CITATION = JHEPA,0704,080;%%

\bibitem{ref:phenomenology}
D.~H.~Rischke, M.~I.~Gorenstein, A.~Schafer, H.~Stoecker,  W.~Greiner,
  %``Nonperturbative effects in the SU(3) gluon plasma,''
  Phys.\ Lett.\  B {\bf 278} (1992) 19;
  %%CITATION = PHLTA,B278,19;%%
A.~Peshier, B.~Kampfer, O.~P.~Pavlenko,  G.~Soff,
  %``A Massive Quasiparticle Model Of The SU(3) Gluon Plasma,''
  Phys.\ Rev.\  D {\bf 54}, 2399 (1996);
  %%CITATION = PHRVA,D54,2399;%%
P.~Levai,  U.~W.~Heinz,
  %``Massive gluons and quarks and the equation of state obtained from SU(3)
  %lattice QCD,''
  Phys.\ Rev.\  C {\bf 57}, 1879 (1998)
  [arXiv:hep-ph/9710463];
  %%CITATION = PHRVA,C57,1879;%%
V.~M.~Bannur,
  %``Quasi-particle model (qQGP) for strongly coupled plasma (SCQGP),''
  [arXiv:0807.2092].
  %%CITATION = ARXIV:0807.2092;%%

\bibitem{viscosity}
  D.~Teaney,
  %``Effect of shear viscosity on spectra, elliptic flow, and Hanbury
  %Brown-Twiss radii,''
  Phys.\ Rev.\  C {\bf 68}, 034913 (2003)
  [arXiv:nucl-th/0301099];
  %%CITATION = PHRVA,C68,034913;%%
  A.~Nakamura, S.~Sakai,
  %``Transport coefficients of gluon plasma,''
  Phys.\ Rev.\ Lett.\  {\bf 94}, 072305 (2005)
  [hep-lat/0406009];
  %%CITATION = PRLTA,94,072305;%%
  H.~B.~Meyer,
  %``A calculation of the shear viscosity in SU(3) gluodynamics,''
  Phys.\ Rev.\  D {\bf 76}, 101701(R) (2007)
  [arXiv:0704.1801].
  %%CITATION = PHRVA,D76,101701;%%

\bibitem{chernodub}
M.~N.~Chernodub, V.~I.~Zakharov,
%``Magnetic component of Yang-Mills plasma,''
Phys.\ Rev.\ Lett.\  {\bf 98}, 082002 (2007) [hep-ph/0611228].
%%CITATION = PRLTA,98,082002;%%

\bibitem{Shuryak:Liao}
J.~Liao, E.~Shuryak,
%``Strongly coupled plasma with electric and magnetic charges,''
Phys.\ Rev.\  C {\bf 75}, 054907 (2007) [hep-ph/0611131].
%%CITATION = PHRVA,C75,054907;%%

\bibitem{chris}
  P.~Giovannangeli, C.~P.~Korthals Altes,
  %``'t Hooft and Wilson loop ratios in the QCD plasma,''
  { Nucl.\ Phys.} {\bf B608}, 203 (2001) [hep-ph/0102022];
  %%CITATION = NUPHA,B608,203;%%
  C.~P.~Korthals Altes,
  in {``Minneapolis 2006, Continuous advances in QCD''}, p.~26
  [hep-ph/0607154];
  %%CITATION = HEP-PH/0607154;%%
M.~Baker,
  %``Understanding Confinement From Deconfinement,''
  Phys.\ Rev.\  D {\bf 78}, 014009 (2008)
  [arXiv:0711.4861 [hep-ph]].
  %%CITATION = PHRVA,D78,014009;%%

\bibitem{greensite}
J.~Greensite,
%``The confinement problem in lattice gauge theory,''
Prog.\ Part.\ Nucl.\ Phys.\  {\bf 51}, 1 (2003) [hep-lat/0301023].
%%CITATION = PPNPD,51,1;%%

\bibitem{DualSuperconductor}
Y.~Nambu,
%``Strings, monopoles, and gauge fields,''
Phys.\ Rev.\  D {\bf 10}, 4262 (1974);
%%CITATION = PHRVA,D10,4262;%%
G.~'t~Hooft, in {High Energy Physics}, ed. A. Zichichi,
EPS International Conference, Palermo (1975);
S.~Mandelstam, { Phys.\ Rept.}  {\bf 23}, 245 (1976);
%%CITATION = PRPLC,23,245;%%
 T.~Suzuki and I.~Yotsuyanagi,
  %``A possible evidence for Abelian dominance in quark confinement,''
  Phys.\ Rev.\  D {\bf 42}, 4257 (1990).
  %%CITATION = PHRVA,D42,4257;%%

\bibitem{ref:review:monopoles}
M. N. Chernodub,  M. I. Polikarpov, in ``Confinement, duality, and nonperturbative aspects of QCD'',
Ed. by Pierre Van Baal (New York, Plenum Press, 1998), hep-th/9710205;
%%CITATION = HEP-TH/9710205;%%
M.~N.~Chernodub, F.~V.~Gubarev, M.~I.~Polikarpov,  A.~I.~Veselov,
%``Monopoles in the Abelian projection of gluodynamics,''
Prog.\ Theor.\ Phys.\ Suppl.\  {\bf 131}, 309 (1998) [arXiv:hep-lat/9802036].
%%CITATION = PTPSA,131,309;%%.

\bibitem{ref:chains}
   J.~Ambjorn, J.~Giedt, J.~Greensite,
  %``Vortex structure vs monopole dominance in Abelian projected gauge
  %theory,''
  { JHEP} {\bf 0002}, 033 (2000)
  [hep-lat/9907021];
  %%CITATION = JHEPA,0002,033;%%
  V.~I.~Zakharov,
  %``Dual string from lattice Yang-Mills theory,''
  { AIP Conf.\ Proc. } {\bf 756}, 182 (2005)
  [hep-ph/0501011].
  %%CITATION = APCPC,756,182;%%

\bibitem{ref:net:analytical}
J.~M.~Cornwall,
  %``Nexus solitons in the center vortex picture of {QCD},''
  Phys.\ Rev.\  D {\bf 58}, 105028 (1998)
  [arXiv:hep-th/9806007].
  %%CITATION = PHRVA,D58,105028;%%

\bibitem{ref:net:numerical}
B.~L.~G.~Bakker, A.~I.~Veselov,  M.~A.~Zubkov,
  %``Central dominance and the confinement mechanism in gluodynamics,''
  Phys.\ Lett.\  B {\bf 471}, 214 (1999)
  [arXiv:hep-lat/9902010].
  %%CITATION = PHLTA,B471,214;%%

\bibitem{ref:monvort}
M.~A.~C.~Kneipp,
  %``Z(k) string fluxes and monopole confinement in non-Abelian theories,''
  Phys.\ Rev.\  D {\bf 68}, 045009 (2003)
  [hep-th/0211049];
  %%CITATION = PHRVA,D68,045009;%%
  %``Color superconductivity, Z(N) flux tubes and monopole confinement in
  %deformed N = 2* super Yang-Mills theories,''
  Phys.\ Rev.\ D {\bf 69}, 045007 (2004)
  [arXiv:hep-th/0308086];
  %%CITATION = PHRVA,D69,045007;%%
R.~Auzzi, S.~Bolognesi, J.~Evslin, K.~Konishi,
  %``Nonabelian monopoles and the vortices that confine them,''
  Nucl.\ Phys.\  {\bf B686}, 119 (2004)
  [hep-th/0312233];
  %%CITATION = NUPHA,B686,119;%%
D.~Tong,
  %``Monopoles in the Higgs phase,''
  Phys.\ Rev.\  D {\bf 69}, 065003 (2004)
  [hep-th/0307302];
  %%CITATION = PHRVA,D69,065003;%%
A.~Gorsky, M.~Shif\-man, A.~Yung,
  %``Non-Abelian Meissner effect in Yang-Mills theories at weak coupling,''
  {\it ibid.} {\bf 71}, 045010 (2005)
  [hep-th/0412082];
  %%CITATION = PHRVA,D71,045010;%%
Y.~Isozumi, M.~Nitta, K.~Ohashi and N.~Sakai,
  %``All exact solutions of a 1/4 Bogomol'nyi-Prasad-Sommerfield equation,''
  Phys.\ Rev.\  D {\bf 71}, 065018 (2005)
  [arXiv:hep-th/0405129];
  %%CITATION = PHRVA,D71,065018;%%
M.~Eto, Y.~Isozumi, M.~Nitta, K.~Ohashi and N.~Sakai,
  %``Instantons in the Higgs phase,''
  Phys.\ Rev.\  D {\bf 72}, 025011 (2005)
  [arXiv:hep-th/0412048];
  %%CITATION = PHRVA,D72,025011;%%
%M.~Eto, Y.~Isozumi, M.~Nitta, K.~Ohashi and N.~Sakai,
  %``Solitons in the Higgs phase: The moduli matrix approach,''
  J.\ Phys.\ A  {\bf 39}, R315 (2006)
  [arXiv:hep-th/0602170];
  %%CITATION = JPAGB,A39,R315;%%
M.~Shifman, A.~Yung,
  %``Supersymmetric Solitons and How They Help Us Understand Non-Abelian   Gauge
  %Theories,''
  Rev.\ Mod.\ Phys.\  {\bf 79}, 1139 (2007)
  [hep-th/0703267].
  %%CITATION = RMPHA,79,1139;%%

\bibitem{ref:monvort:ab}
M.N.~Cherno\-dub, R.~Feldmann, E.-M.~Il\-gen\-fritz, A.~Schil\-ler, { Phys. Lett.} {\bf B605}, 161 (2005) [hep-lat/0406015].
%%CITATION = PHLTA,B605,161;%%

\bibitem{ref:review:MSU}
M.~N.~Chernodub, V.~I.~Zakharov,
``Monopoles and vortices in Yang-Mills plasma'',
to appear in Phys. Atom. Nucl. [arXiv:0806.2874].
%%CITATION = ARXIV:0806.2874;%%

\bibitem{vz}
V.~I.~Zakharov,
%``From confining fields on the lattice to higher dimensions in the continuum,''
Braz.\ J.\ Phys.\  {\bf 37}, 165 (2007) [hep-ph/0612342];
%%CITATION = BJPHE,37,165;%%
%V.~I.~Zakharov,
%``Dual string from lattice Yang-Mills theory,''
AIP Conf.\ Proc.\  {\bf 756}, 182 (2005) [hep-ph/0501011].
%%CITATION = APCPC,756,182;%%

\bibitem{Langfeld:1998cz}
  K.~Langfeld, O.~Tennert, M.~Engelhardt,  H.~Reinhardt,
  %``Center vortices of Yang-Mills theory at finite temperatures,''
  Phys.\ Lett.\  B {\bf 452}, 301 (1999) [hep-lat/9805002].
  %%CITATION = PHLTA,B452,301;%%

\bibitem{zubkov}
M.~N.~Chernodub, M.~I.~Polikarpov, A.~I.~Veselov, M.~A.~Zubkov,
%``Aharonov-Bohm effect, center monopoles and center vortices in SU(2)
%lattice gluodynamics,''
Nucl.\ Phys.\ Proc.\ Suppl.\  {\bf 73}, 575 (1999)
[hep-lat/9809158].
%%CITATION = NUPHZ,73,575;%%

\bibitem{jeff:finite}
R.~Bertle, M.~Faber, J.~Greensite,  S.~Olejnik,
  %``The structure of projected center vortices in lattice gauge theory,''
  JHEP {\bf 9903}, 019 (1999)
  [arXiv:hep-lat/9903023].
  %%CITATION = JHEPA,9903,019;%%

\bibitem{langfeld2}
M.~Engelhardt, K.~Langfeld, H.~Reinhardt, O.~Tennert,
  %``Deconfinement in SU(2) Yang-Mills theory as a center vortex percolation
  %transition,''
  Phys.\ Rev.\  D {\bf 61}, 054504 (2000) [hep-lat/9904004].
  %%CITATION = PHRVA,D61,054504;%%

\bibitem{ref:Lattice}
 M.~N.~Chernodub, K.~Ishiguro, A.~Nakamura, T.~Sekido, T.~Suzuki, V.~I.~Zakharov,
  %``Topological defects and equation of state of gluon plasma,''
  PoS {\bf LAT2007}, 174 (2007)
  [arXiv:0710.2547].
  %%CITATION = POSCI,LAT2007,174;%%

\bibitem{Engels:1988ph}
J.~Engels, F.~Karsch, H.~Satz, I.~Montvay,
  %``Gauge Field Thermodynamics For The SU(2) Yang-Mills System,''
  Nucl.\ Phys.\  B {\bf 205}, 545 (1982);
  %%CITATION = NUPHA,B205,545;%%
J.~Engels, J.~Fingberg, K.~Redlich, H.~Satz, M.~Weber,
%``THE ONSET OF DECONFINEMENT IN SU(2) LATTICE GAUGE THEORY,''
Z.\ Phys.\  C {\bf 42}, 341 (1989);
%%CITATION = ZEPYA,C42,341;%%
J.~Engels, J.~Fingberg, F.~Karsch, D.~Miller, M.~Weber,
  %``Nonperturbative thermodynamics of SU(N) gauge theories,''
  Phys.\ Lett.\  B {\bf 252}, 625 (1990);
  %%CITATION = PHLTA,B252,625;%%
J.~Engels, F.~Karsch, K.~Redlich,
  %``Scaling Properties Of The Energy Density In SU(2) Lattice Gauge Theory,''
  Nucl.\ Phys.\  B {\bf 435}, 295 (1995)
  [hep-lat/9408009].
  %%CITATION = NUPHA,B435,295;%%

\bibitem{ref:Karsch:SU3}
G.~Boyd, J.~Engels, F.~Karsch, E.~Laermann, C.~Legeland, M.~Lutgemeier, B.~Petersson,
%``Equation of state for the SU(3) gauge theory,''
Phys.\ Rev.\ Lett.\  {\bf 75}, 4169 (1995) [hep-lat/9506025];
%%CITATION = PRLTA,75,4169;%%
%G.~Boyd, J.~Engels, F.~Karsch, E.~Laermann, C.~Legeland, M.~Lutgemeier and B.~Petersson,
%``Thermodynamics of SU(3) Lattice Gauge Theory,''
Nucl.\ Phys.\  B {\bf 469}, 419 (1996) [hep-lat/9602007].
%%CITATION = NUPHA,B469,419;%%

\bibitem{Engels:1989fz}
  J.~Engels, J.~Fingberg, M.~Weber,
  %``FINITE SIZE SCALING ANALYSIS OF SU(2) LATTICE GAUGE THEORY IN
  %(3+1)-DIMENSIONS,''
  Nucl.\ Phys.\  B {\bf 332}, 737 (1990).
  %%CITATION = NUPHA,B332,737;%%

\bibitem{Fingberg:1992ju}
  J.~Fingberg, U.~M.~Heller, F.~Karsch,
  %``Scaling And Asymptotic Scaling In The SU(2) Gauge Theory,''
  Nucl.\ Phys.\  B {\bf 392}, 493 (1993)
  [hep-lat/9208012];
  %%CITATION = NUPHA,B392,493;%%
M.~J.~Teper,
  %``Glueball masses and other physical properties of SU(N) gauge theories  in D
  %= 3+1: A review of lattice results for theorists,''
  hep-th/9812187.
  %%CITATION = HEP-TH/9812187;%%

\bibitem{ref:beta:interpolation}
J.~C.~R.~Bloch, A.~Cucchieri, K.~Langfeld, T.~Mendes,
  %``Propagators and running coupling from SU(2) lattice gauge theory,''
  Nucl.\ Phys.\  B {\bf 687}, 76 (2004)
  [hep-lat/0312036].
  %%CITATION = NUPHA,B687,76;%%

\bibitem{ref:asymmetry}
M.~N.~Chernodub and E.~M.~Ilgenfritz,
  %``Electric-magnetic asymmetry of the A^2 condensate and the phases of
  %Yang-Mills theory,''
  Phys.\ Rev.\  D {\bf 78}, 034036 (2008)
  [arXiv:0805.3714 [hep-lat]].
  %%CITATION = PHRVA,D78,034036;%%

\bibitem{ref:removed:vortices}
P.~de Forcrand, M.~D'Elia,
  %``On the relevance of center vortices to QCD,''
  Phys.\ Rev.\ Lett.\  {\bf 82}, 4582 (1999) [hep-lat/9901020].
  %%CITATION = PRLTA,82,4582;%%

\bibitem{ref:center:vortex}
  L.~Del Debbio, M.~Faber, J.~Greensite, S.~Olejnik,
  %``Center dominance and Z(2) vortices in SU(2) lattice gauge theory,''
  Phys.\ Rev.\  D {\bf 55}, 2298 (1997) [hep-lat/9610005];
  %%CITATION = PHRVA,D55,2298;%%
L.~Del Debbio, M.~Faber, J.~Giedt, J.~Greensite, S.~Olejnik,
  %``Detection of center vortices in the lattice Yang-Mills vacuum,''
  Phys.\ Rev.\  D {\bf 58}, 094501 (1998) [hep-lat/9801027].
  %%CITATION = PHRVA,D58,094501;%%

\bibitem{ref:cautionary}
  V.~G.~Bornyakov, D.~A.~Komarov, M.~I.~Polikarpov,
  %``P-vortices and drama of Gribov copies,''
  Phys.\ Lett.\  B {\bf 497}, 151 (2001)
  [hep-lat/0009035].
  %%CITATION = PHLTA,B497,151;%%

\bibitem{ref:Kapusta}
J.~I.~Kapusta, ``Finite-Temperature Field Theory''
(Cambridge University Press, Cambridge, 1989).

\bibitem{ref:Italians}
  A.~D'Alessandro, M.~D'Elia,
  %``Magnetic monopoles in the high temperature phase of Yang-Mills theories,''
 { Nucl.\ Phys.}   {\bf B799}, 241 (2008)
  [arXiv:0711.1266].
  %%CITATION = NUPHA,B799,241;%

\bibitem{Shuryak}
 J.~Liao, E.~Shuryak,
%  {``Magnetic Component of Quark-Gluon Plasma is also a  Liquid!,''},
   arXiv:0804.0255.
  %%CITATION = ARXIV:0804.0255;%%

\bibitem{ref:blocking}
  M.~N.~Chernodub, K.~Ishiguro, T.~Suzuki,
  %``Blocking of lattice monopoles from the continuum in hot lattice
  %gluodynamics,''
  { JHEP} {\bf 0309} 027, (2003) [hep-lat/0204003].
  %%CITATION = JHEPA,0309,027;%%

\bibitem{ref:fine:tuning}
F.~V.~Gubarev, A.~V.~Kovalenko, M.~I.~Polikarpov, S.~N.~Syritsyn, V.~I.~Zakharov,
  %``Fine tuned vortices in lattice SU(2) gluodynamics,''
  Phys.\ Lett.\  B {\bf 574}, 136 (2003) [hep-lat/0212003];
  %%CITATION = PHLTA,B574,136;%%
A.~V.~Kovalenko, M.~I.~Polikarpov, S.~N.~Syritsyn, V.~I.~Zakharov,
  %``Properties of P-vortex and monopole clusters in lattice SU(2) gauge
  %theory,''
  Phys.\ Rev.\  D {\bf 71}, 054511 (2005)  [hep-lat/0402017].
  %%CITATION = PHRVA,D71,054511;%%

\bibitem{ref:nonAbelian:vortices}
A.~Gorsky, V.~Zakharov,
  %``Magnetic strings in Lattice QCD as Nonabelian Vortices,''
  Phys.\ Rev.\  D {\bf 77}, 045017 (2008)
  [arXiv:0707.1284].
  %%CITATION = PHRVA,D77,045017;%%

\bibitem{ref:Larry}
L.~McLerran,
  %``Gluon Saturation and the Formation Stage of Heavy Ion Collisions,''
  arXiv:0807.4095;
  %%CITATION = ARXIV:0807.4095;%%
A.~Dumitru, F.~Gelis, L.~McLerran and R.~Venugopalan,
  %``Glasma flux tubes and the near side ridge phenomenon at RHIC,''
  Nucl.\ Phys.\  A {\bf 810}, 91 (2008)
  [arXiv:0804.3858 [hep-ph]].
  %%CITATION = NUPHA,A810,91;%%

\bibitem{ref:preparation}
M.N.Chernodub, V.I.Zakharov, in preparation.

\end{thebibliography}
\end{document}